\documentclass[%
reprint,
prd,f
amsmath,amssymb,
aps,
lengthcheck,
floatfix,
]{revtex4}

\usepackage{graphicx}% Include figure files
\usepackage{dcolumn}% Align table columns on decimal point
\usepackage{bm}% bold math
\usepackage{hyperref}% add hypertext capabilities
\usepackage[utf8]{inputenc}
\usepackage{multirow}
\usepackage{soul}
\usepackage{float}
\usepackage{graphicx}
\usepackage{times}
\usepackage[normalem]{ulem}
\usepackage{color}
\usepackage{cellspace}
\usepackage[usenames,dvipsnames]{xcolor}

\usepackage{lipsum}% http://ctan.org/pkg/lipsum

\newcommand{\be}{\begin{equation}}
	\newcommand{\ee}{\end{equation}}
\newcommand{\bea}{\begin{eqnarray}}
	\newcommand{\eea}{\end{eqnarray}}

\newcommand{\comment}[1]{}
\renewcommand\sout{\bgroup \color{red} \ULdepth=-.5ex \ULset}
\def\simge{\mathrel{\rlap{\raise 0.511ex
			\hbox{$>$}}{\lower 0.511ex \hbox{$\sim$}}}}
\def\simle{\mathrel{\rlap{\raise 0.511ex
			\hbox{$<$}}{\lower 0.511ex \hbox{$\sim$}}}}

\DeclareUnicodeCharacter{2212}{-}
\usepackage[none]{hyphenat}

\begin{document}
	
	% \setcounter{page}{1}
	% \vspace*{0.3 true in}
	\title{Systematic analysis of the impacts of symmetry energy parameters on neutron star properties}

	\author{\href{https://orcid.org/0000-0003-0103-5590}N. K. Patra$^{1}$\includegraphics[scale=0.06]{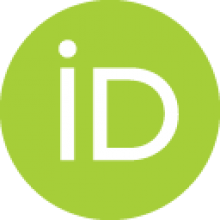}}
	\email{nareshkumarpatra3@gmail.com}
	\author{\href{https://orcid.org/0000-0002-0812-2702}Anagh Venneti $^{2}$\includegraphics[scale=0.06]{Orcid-ID.png}}
	
	\author{\href{https://orcid.org/0000-0003-3308-2615}Sk Md Adil Imam$^{3,4}$\includegraphics[scale=0.06]{Orcid-ID.png}}
	
	\author{\href{https://orcid.org/0000-0003-1274-5846}Arunava Mukherjee$^{3,4}$\includegraphics[scale=0.06]{Orcid-ID.png}}
	\email{arunava.mukherjee@saha.ac.in}
	
	\author{\href{https://orcid.org/0000-0001-5032-9435}B. K. Agrawal$^{3,4}$\includegraphics[scale=0.06]{Orcid-ID.png}
	}
	\email{bijay.agrawal@saha.ac.in}

	\affiliation{$^1$Department of Physics, BITS-Pilani, K. K. Birla Goa Campus, Goa 403726, India}
	\affiliation{$^2$Department of Physics, BITS-Pilani, Hyderabad Campus, Hyderabad - 500078, India }
	\affiliation{$^3$Saha Institute of Nuclear Physics, 1/AF 
		Bidhannagar, Kolkata 700064, India}  
	\affiliation{$^4$Homi Bhabha National Institute, Anushakti Nagar, Mumbai 400094, India.}

	\date{\today}
	
	\begin{abstract} 
		
		The impacts of various symmetry energy parameters on the properties of
		neutron stars (NSs)  have been recently investigated, and the outcomes are
		at variance, as summarized in Table III of Phys. Rev. D 106, 063005 (2022).
		We have systematically analyzed the correlations of slope and curvature
		parameters of symmetry energy at the saturation density ($\rho_0=0.16
		\text{fm}^{-3}$) with the tidal deformability and stellar radius of
		non-spinning neutron stars in the mass range of $1.2 - 1.6 M_\odot$
		using a large set of minimally constrained equations of state (EoSs).
		The EoSs at low densities correspond to the nucleonic matter
		and are  constrained   by empirical ranges of a few low-order nuclear matter
		parameters from the finite nuclei data and the pure neutron matter EoS
		from chiral effective field theory.  The EoSs at high densities ($\rho >
		1.5 - 2\rho_0$)  are obtained by a parametric form for the speed of sound that
		satisfies the causality condition.  Several factors  affecting the
		correlations between the NS properties and the individual symmetry energy parameters usually considered in the literature are explored. These correlations are    quite sensitive
		to the choice of the distributions of symmetry energy  parameters and
		their interdependence.
		But,  variations of  NS properties with the pressure of $\beta -$ equilibrated matter at twice  the saturation density remain quite robust which 
		maybe due to the fact that the pressure depends on the combination of
		multiple nuclear matter parameters that describe the symmetric nuclear
		matter as well as the density dependence of the symmetry energy.
		Our results are practically insensitive to the behavior of EoS at high
		densities.
		
	\end{abstract}
	
	%\pacs{21.30.Fe, 21.65.Cd, 21.65.Mn, 21.65.Ef}
	%\keyword{Neutron Star, Equation of State, Symmetry energy ,  Bayesian Parameter Estimation}
	
	\maketitle
	\section{Introduction}
	The equations of state (EoSs)  of $\beta$-equilibrated charge neutral matter and their connections to the properties of neutron stars (NSs) have been studied  for the last several decades~\cite{Oppenheimer:1939ne, Tolman:1939jz, Glendenning:1992dr}. The precise knowledge of the properties of NS and the data on Heavy ion collisions may constrain the behavior of EoSs at supra-saturation densities \cite{Huth:2021bsp}. The radius and tidal deformability from a population of neutron stars over a wide range of mass ($1 - 3 M_\odot$) would probe the EoS at densities upto a few times ($\sim 2 - 8$) of the saturation density $\rho_0 \approx 0.16$fm$^{-3}$ encountered at the center of finite nuclei. The tidal deformability parameter of NS, which encodes the information about the EoS has been inferred for the first time from a gravitational wave event GW170817 observed by the advanced-LIGO~\cite{LIGOScientific:2014pky} and advanced-Virgo detectors~\cite{VIRGO:2014yos} from a binary neutron star (BNS) merger with a total mass of the system, $2.74_{-0.01}^{+0.04}M_\odot$ ~\cite{Abbott18a, Abbott2019}. Another subsequent event GW190425, likely originating from the coalescence of BNSs was observed~\cite{Abbott2020}. The BNS signals emitted from coalescing neutron stars are likely to be observed more frequently in the upcoming runs of LIGO-Virgo-KAGRA and the future detectors, e.g., Einstein Telescope~\cite{Punturo:2010zz} and Cosmic Explorer~\cite{Reitze:2019iox}. The unprecedented constraints on the EoS promised by gravitational wave astronomy through the detailed analysis of gravitational wave parameter estimation have triggered many theoretical investigations of the NS properties ~\cite{GW170817, Malik2018, De18, Fattoyev2018a, Forbes_etal_withAM2019, Landry2019, Piekarewicz2019, Biswas2020, Abbott2020, Thi2021}. 
	Recently, two different groups of Neutron star Interior Composition Explorer (NICER) X-ray telescopes provided neutron star's mass and radius simultaneously for PSR J0030+0451 with $R=13.02^{+1.24}_{-1.06}$km for mass $1.44^{+0.15}_{-0.14} M_\odot$ \cite{Miller2019} and $R=12.71^{+1.14}_{-1.19}$km  for mass $1.34^{+0.15}_{-0.16} M_\odot$\cite{Riley2019}, which are the complementary constraints on the EoS. For heavier pulsar PSR J0740+6620, $R=13.7^{+2.6}_{-1.5}$km with mass $2.08 \pm 0.07 M_\odot$  \cite{Miller:2021qha} and $R=12.39^{+1.30}_{-0.98}$km with mass $2.072^{+0.067}_{-0.066} M_\odot$ \cite{Riley:2021pdl} were reported. Current observational lower bound on the maximum NS mass is $M_{\rm max}= 2.35\pm 0.17 M_\odot$ for the black-widow pulsar PSR J0952-0607 \cite{Romani:2022jhd} that exceeds any previous measurements, including $M_{\rm max} = 2.27^{+0.17}_{-0.15} M_\odot$ for PSR J2215-5135\cite{Linares:2018ppq}. If the observational bounds are reliable, stiffer EoSs are required to support the NS with a mass higher than $2 M_\odot$.
	
	The present lower bound on $M_{\rm max}$ suggests that the density at the center of the star with canonical mass $1.4 M_\odot$ may lie in the range $\sim 2-3\rho_0$\cite{Hebeler:2013nza, Li:2005sr}. 
	The behavior of EoSs around $\rho_0$ may be important in determining the properties of such NSs. It has been shown that the radius of a neutron star with its mass in the range of $1-1.4 M_\odot$ is strongly correlated with the pressure for $\beta$-equilibrated matter at the densities $1-2\rho_0$ \cite{Lattimer:2000nx}. Similar analyses have been extended to the tidal deformability, which is also found to be strongly correlated with pressure at $2\rho_0$ \cite{Tsang:2019vxn, Tsang:2020lmb, Patra:2022yqc}. The EoS for $\beta$-equilibrated matter can be decomposed into that for symmetric nuclear matter and density-dependent symmetry energy. It may be important to constrain them individually. 
	{Recently, it is shown in Ref. \cite{Imam:2021dbe} that the accurate knowledge of the equation of state of $\beta$-equilibrated matter may not be resolved appropriately into its two main components, symmetric nuclear matter, and density-dependent symmetry energy. Similar conclusions are also drawn in references \cite{deTovar:2021sjo, Mondal:2021vzt}.}

	There have been several attempts to study the correlations of radius and tidal deformability of a neutron star with individual nuclear matter parameters which determine the density-dependence of symmetry energy~\cite{Alam:2016cli, Carson:2018xri, Malik2018, Tsang:2019vxn, Guven:2020dok, Malik:2020vwo, Tsang:2020lmb, Malik_book, Reed:2021nqk, Pradhan:2022vdf, Pradhan:2022txg, Ghosh:2022lam, Ghosh:2021bvw, Beznogov:2022rri}. The nuclear matter parameters, often drawn randomly from uncorrelated uniform or Gaussian distributions, are found to be weakly correlated with the NS properties. The distributions of nuclear matter parameters obtained by fitting the experimental data on finite nuclei properties within the mean-field models are somewhat better correlated with the NS properties. Lately, the impact on the NS properties due to constraints on the low-density EoS imposed by the finite nuclei data have been investigated using more than four hundred mean-field models derived from the non-relativistic Skyrme interactions and relativistic lagrangians which describe the interactions of nucleons through $\sigma$, $\omega$, and $\rho$ mesons \cite{Carlson:2022nfb}. The group of models that describe very well the properties of symmetric and asymmetric finite nuclei yields a stronger correlation between the tidal deformability for the NS with canonical $1.4 M_\odot$ mass star and the slope of the symmetry energy at the saturation density ($\rho_0$). Several factors that affect the correlations of NS properties with the nuclear matter parameters are also summarized in Table III of Ref.~\cite{Kunjipurayil:2022zah}.
	
	We have considered a large set of minimally constrained EoSs for the NS matter in the present work to examine in detail the correlations between the properties of a neutron star in the mass range $1.2 - 1.6 M_{\odot}$ and the parameters that govern the density dependence of symmetry energy. The EoSs at low densities corresponds to the nucleonic matter in $\beta$-equilibrium and are described by the nuclear matter parameters evaluated at $\rho_0$. These EoSs are constrained by empirical values of the low-order nuclear matter parameters determined by the experimental data on the bulk properties of finite nuclei together with the pure neutron matter (PNM) EoS from a precise next-to-next-to-next-to-leading-order (N$^{3}$LO) calculation in chiral effective field theory. The composition of NS matter at high density ($\rho>2\rho_0$) is not very well known due to the possibility of the appearance of various new degrees of freedom such as hyperons, kaons, and quarks \cite{Chatziioannou:2015uea, Chatterjee:2015pua, Stone:2019blq}. Beyond a transition density ($\rho_{\rm tr}$), taken to be $1.5 - 2\rho_0$, the EoSs are constructed simply by imposing the causality condition on the speed of sound and are independent of compositions of NS matter. 
	The posterior distributions of nuclear matter parameters that describe the low-density EoSs are obtained within a Bayesian approach with minimal constraints. These  constraints introduce correlations among the nuclear matter parameters. The joint posterior or correlated distribution of the nuclear matter parameters is employed to study the sensitivity of NS
	properties to  the  parameters that govern the density dependence of the symmetry energy. The calculations are also performed for uncorrelated uniform and Gaussian distributions of the nuclear matter parameters obtained by their marginalized posterior distribution. The influence of the various correlations considered due to a few other factors usually encountered in the literature are investigated. 
	
	The paper is organized as follows.  We briefly outline our methodology in Sec.\ref{methe}. The results for the correlations of NS properties with various symmetry energy parameters at $\rho_0$ and the pressure for $\beta$-equilibrated matter at $2\rho_0$ are discussed in detail in Sec. \ref{results}. The summary and outlook are presented in Sec. \ref{summary}. 
	
	%%%%%%%%%%%%%%%%%%%%%%%%%%%%%%%%%%%%%%%%%%
	\section{Methodology} 
	\label{methe}
	We discuss in brief the construction of the equation of state (EoS) at low and high densities. The low-density EoS below 2$\rho_0$ is obtained using Taylor expansion, and the high-density EoS is constructed keeping a check on the speed of sound so that the causality is not violated. 
	
	\subsection{{EoS at Low density}}\label{L-eos}
	We express energy per nucleon at a given density $\rho$ and asymmetry $\delta$  using  parabolic approximation as,
	\bea
	E(\rho, \delta) &=&  E(\rho,0)+E_{\rm sym}(\rho)\delta^2
	+..., \label{eq:EoS}
	\eea
	where $\delta = \frac{\rho_n -\rho_p}{\rho}$
	is determined  using the $\beta$-equilibrium and the charge neutrality conditions. The energy per nucleon for the symmetric nuclear matter, $E(\rho,0)$ and the density-dependent symmetry energy, $E_{\rm sym}(\rho)$ are expanded around  $\rho_0$ using individual nuclear matter parameters as \cite{Chen:2005ti,Chen:2009wv,Newton:2014iha,Margueron:2017eqc,Margueron:2018eob},
	\bea
	E(\rho, 0)&=&	e_0+\frac{1}{2}K_0\left (\frac{\rho-\rho_0}{3\rho_0}\right )^2
	+\frac{1}{6} Q_0 \left(\frac{\rho-\rho_0}{3\rho_0}\right )^3, \label{eq:SNM_T} \\ 
	E_{\rm sym}(\rho) &=&	J_0 + L_0\left (\frac{\rho-\rho_0}{3\rho_0}\right )
	+\frac{1}{2}K_{\rm sym,0}\left (\frac{\rho-\rho_0}{3\rho_0}\right )^2 \nonumber \\  
	&+& \frac{1}{6}Q_{\rm sym,0}\left (\frac{\rho-\rho_0}{3\rho_0}\right )^3. \label{eq:sym_T}  
	\eea 
	In Eqs. (\ref{eq:SNM_T}) and (\ref{eq:sym_T}), $e_0 $ is the binding energy per nucleon, $K_0$ is the incompressibility coefficient, $J_0$ is the symmetry energy coefficient, its slope parameter $L_0$, $K_{\rm sym,0}$ the symmetry energy curvature parameter, $Q_0 [Q_{\rm sym,0}]$ is related to third-order density derivatives of  $E (\rho,0) $  $[E_{\rm sym}(\rho)]$.
	
	\subsection{EoS at High density}\label{H-eos}
	We impose the causality condition on the speed of sound to construct the EoS beyond the transition density ($\rho_{\rm tr}$), which is taken to be $1.5-2\rho_0$. The high-density part of the EoS ($\rho>\rho_{\rm tr}$) joins smoothly to the one at the low density such that the velocity of the sound never exceeds the velocity of light and asymptotically approaches the conformal limit ($c_s^2$ = $\frac{1}{3}c^2$). The velocity of sound for $\rho>\rho_{\rm tr}$ is given as\cite{Tews:2018kmu},   
	\bea
	\frac{c_s^2}{c^2} &=& \frac{1}{3} - c_1 exp\left[{-\frac{(\rho-c_2)^2}{n_b^2}}\right] + h_p exp\left[-\frac{(\rho-n_p)^2}{w_p^2}\right]\nonumber\\
	&& \left[1 + erf(s_p \frac{\rho-n_p}{w_p})\right]. \label{eq-vs}
	\eea 
	where the peak height $h_p$ determines the maximum speed of sound, the position $n_p$ determines the density around which it happens, the width of the curve controls by $w_p$ and $n_b$, and the shape or skewness parameter $s_p$. For a given value of $n_b$, the parameters $c_1$ and $c_2$ are determined by the continuity of the speed of sound and its derivative at the transition density $\rho_{\rm tr}$. The values of $n_b$, $h_p$, $w_p$, and $n_p$ are drawn from the uniform distribution with ranges in between 0.01-3.0 fm$^{-3}$,  0.0-0.9, 0.1-5.0 fm$^{-3}$, and ($\rho_{\rm tr}$ + 0.08) - 5.0 fm$^{-3}$, respectively \cite{Tews:2018kmu}. We have taken $s_p$ equal to zero throughout our calculations as it does not have much effect on the stiffness of EoS. 
	
	We construct the high-density equation of state starting from transition density ($\rho_{\rm tr}$), where the energy density ($\epsilon(\rho_{\rm tr})$), the pressure (P($\rho_{\rm tr}$)) and the derivative of energy density ($\epsilon^\prime(\rho_{\rm tr})$) are known. {The successive values of $\epsilon$ and P are obtained by assuming a step size  \(\Delta \rho=0.001\) fm$^{-3}$ as follows,} 
	\bea
	\rho_{i+1} &=& \rho_i + \Delta \rho \label{eq-rhoE},\\
	\epsilon_{i+1} &=& \epsilon_i + \Delta \epsilon \nonumber\\
	&=& \epsilon_i + \Delta \rho \frac{\epsilon_i + P_i}{\rho_i}\label{eq-engE},\\
	P_{i+1} &=&P_i + c_s^2(\rho_i) \Delta \epsilon  \label{eq-preE}.              
	\eea
	where, the index i = 0 refers to the transition density $\rho_{\rm tr}$ . Note, in the Eq.(\ref{eq-engE}) $\Delta \epsilon$ has been evaluated using the thermodynamic relation $ P =  \rho \partial \epsilon/{\partial \rho}-\epsilon$ valid at zero temperature. Once the EoS has been generated, the NS properties, such as the radius and tidal deformability as a function of mass, are obtained by solving the Tolman-Oppenheimer-Volkoff (TOV) equations by varying the central density of the star.
	
	\subsection{Bayesian estimation of nuclear matter parameters}\label{BA}
	The detailed  statistical analysis of the parameters of a model for a given set of data can be carried out in the Bayesian approach. It gives the joint posterior distributions of model parameters by which one can study the distributions of given parameters and correlations among the parameters. Based on the Bayes theorem, the joint posterior distribution of the parameters $P(\bm{\theta} |D )$ is given as \cite{Gelman2013},
	
	\begin{equation}
		P(\bm{\theta} |D ) =\frac{{\mathcal L } (D|\bm{\theta}) P(\bm {\theta })}{\mathcal Z},\label{eq:bt}
	\end{equation}
	where D and  $\bm{\theta}$ are the data and set of model parameters, respectively. Here $P(\bm {\theta })$ is the prior for model parameters,  $\mathcal L (D|\bm{\theta})$ is the likelihood function, and $\mathcal Z$ is the evidence.
	The marginalized posterior distribution for a  parameter $\theta_i$
	can be obtained as
	\begin{equation}
		P (\theta_i |D) = \int P(\bm {\theta} |D) \prod_{k\not= i }d\theta_k. \label{eq:mpd}
	\end{equation}
	We use the Gaussian likelihood function, defined as 
	\bea
	{\mathcal L} (D|\bm{\theta})&=&\prod_{j} 
	\frac{1}{\sqrt{2\pi\sigma_{j}^2}}e^{-\frac{1}{2}\left(\frac{d_{j}-m_{j}(\bm{\theta)}}{\sigma_{j}}\right)^2}. 
	\label{eq:likelihood}  
	\eea
	where the index $ j$ runs over all the data, $d_j$ and $m_j$ are the data and corresponding model values, respectively. The $\sigma_j$s correspond to the adopted uncertainties. The posterior distribution of Eq. (\ref{eq:bt}) can be evaluated  by implementing a nested sampling algorithm. We have used the Pymultinest nested sampling  \cite{Buchner2014} in the Bayesian Inference Library  \cite{Ashton2019}.

	\section{Results and Discussion} \label{results}
	
	The EoSs for the asymmetric nuclear matter at low densities are expressed in terms of nuclear matter parameters which are  evaluated at the  saturation density  as outlined in Sec. \ref{L-eos} (Eqs.( \ref{eq:SNM_T}) \& ( \ref{eq:sym_T})).  The joint posterior distribution  of nuclear matter parameters 
	is obtained within a Bayesian approach using minimal constraints. These constraints introduce correlations among the nuclear matter parameters. The EoSs at high density ($\rho>\rho_{\rm tr}$) are constructed by simply imposing the causality condition  on the speed of sound as outlined in Sec. \ref{H-eos}. A large set of EoSs is constructed by appropriately combining the low and high-density parts. These EoSs are employed to evaluate various NS properties, like radius, tidal deformability, and  maximum mass.  We assess the dependence  of radius and  tidal deformability  on the slope and  curvature parameters of the symmetry energy evaluated at $\rho_0$ as well as on the pressure $P(2\rho_0)$ for the $\beta$-equilibrated matter.  The sensitivity of these dependencies to the various factors as follows are analyzed, 
	\begin{itemize}
		\item [(i)] the behavior of the high-density part of the EoS, 
		\item [(ii)] the choice of distributions of nuclear matter parameters, their interdependence, and  uncertainties,
		\item [(iii)] the lower bound on the maximum mass of the stable neutron stars,
		\item [(iv)] the value of the transition density,
		\item [(v)] upper bound on the value of tidal deformability.
	\end{itemize}

	\subsection{Priors and Posterior distributions of NMPs}\label{PD}
	
	\begin{table}[b]
		\caption{\label{tab1}Uniform prior distributions are assumed for all the NMPs except for $e_0$, which is kept fixed to -16.0 MeV. The minimum (min.) and maximum (max.) values of the NMPs are listed in the units of MeV.} 
		\centering
		\begin{ruledtabular}  
			\begin{tabular}{ccccccc}
				%\toprule
				&{$K_0$} & {$Q_0$} & {$J_0$} & {$L_0$} &{$K_{\rm sym,0}$} & {$Q_{\rm sym,0}$} \\ [1.3ex]
				\hline
				min. & 190 & -1200 & 30 & 0 & -500 & -250 \\[1.3ex] 
				max. & 290 & 400 & 35 & 100 & 300 & 1350 \\[1.3ex] 
			\end{tabular}
		\end{ruledtabular}
	\end{table}
	
	\begin{figure*}[t]
		\centering
		\includegraphics[width=0.8\textwidth]{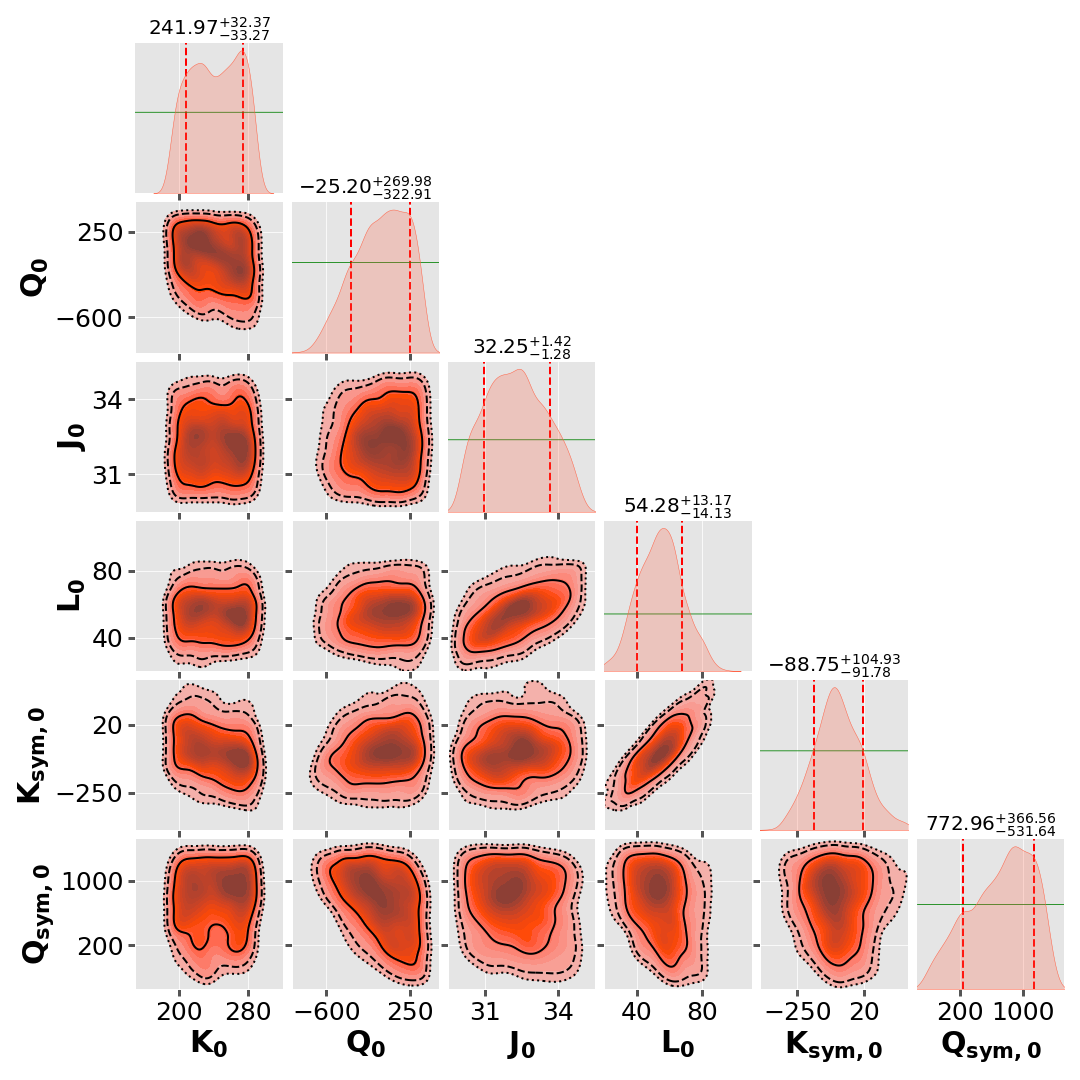}
		\caption{(Color Online)Corner plot for the nuclear matter parameters (in MeV). The one-dimensional marginalized posterior distributions (salmon) and the prior distributions (green) lines are displayed along the diagonal plots. The vertical dashed lines indicate 68$\%$ (1$\sigma$) confidence interval. The confidence ellipses for two-dimensional posterior distributions are plotted with 1$\sigma$, 2$\sigma$, and 3$\sigma$ confidence intervals along the off-diagonal plots.  }\label{fig1}
	\end{figure*}
	
	The posterior distributions of the NMPs are obtained by subjecting the EoSs to a set of minimal constraints, which includes the basic nuclear matter properties at the saturation density and the EoS for the pure neutron matter at low densities from  N$^3$LO calculation in the chiral effective field theory \cite{Hebeler:2013nza}. Only a few low-order NMPs are well constrained, such as the binding energy per nucleon in symmetric nuclear matter $e_0 \sim -16.0 $ MeV and symmetry energy coefficient $J_0=32.5 \pm 2.5 $ MeV from the binding energy of finite nuclei over a wide range of nuclear masses~\cite{Chabanat98, Chabanat97, Malik:2019whk, Mondal:2015tfa, Mondal:2016roo, Sulaksono:2009rn, Essick:2021ezp}. The nuclear matter incompressibility
	coefficient, $K_0=240 \pm 50$ MeV for the symmetric nuclear matter is constrained from the experimental data on the centroid energy of isoscalar giant monopole resonance in a few heavy nuclei~\cite{Garg:2018uam, Agrawal:2005ix}. 
	{The value of  symmetry energy slope parameter has been deduced by the neutron-skin thickness of $^{48}$Ca nucleus by CREX \cite{CREX:2022kgg} and that in $^{208}$Pb by  PREX-2 collaboration \cite{PREX:2021umo}. The PREX-2 data analysis performed by Reed et al. in Ref. \cite{Reed:2021nqk, Reed:2019ezm} places $L_0=106 \pm 37$ MeV. Other studies combining astronomical observations and PREX-2 data is $L_0=53^{+14}_{-15}$ MeV \cite{Essick:2021kjb}. A smaller value of  $L_0=54 \pm 8$ MeV has also been inferred from PREX-2 data \cite{Reinhard:2021utv}. The CREX data predicts $L_0 = 0 - 51$ MeV \cite{TAGAMI2022}.}
	The  remaining nuclear matter parameters, $Q_0$, $K_{\rm sym,0}$ and $Q_{\rm sym,0}$ appearing in Eqs.( \ref{eq:SNM_T}) \& ( \ref{eq:sym_T}) are only weakly constrained ~\cite{Tsang:2020lmb, Ferreira:2021pni, Dutra:2012mb, Dutra:2014qga, Mondal:2017hnh, Patra:2022yqc}. The prior for the binding energy per nucleon is kept fixed to $e_0=-16.0$ MeV throughout. The prior distributions of $J_0$ and $K_0$ are assumed to be uniform with a rather small range, whereas the other higher-order nuclear matter parameters
	correspond to uniform distributions with large ranges. We have listed the assumed prior distributions for each of the nuclear matter parameters in Table \ref{tab1}. The values of $d$ and $\sigma$ in Eq.(\ref{eq:likelihood}) are taken from Ref. \cite{Lattimer:2021emm, Hebeler:2013nza} for the energy per neutron, and we consider a $6\times$N$^3$LO  uncertainty band
	for our calculations. In addition to likelihood and priors, we have imposed a few filters on the nuclear matter parameters: (i) pressure for the $\beta$-equilibrated matter should increase monotonically with density (thermodynamic stability),(ii) symmetry energy is positive semi-definite  and (iii) maximum mass of neutron star must exceed $2M_\odot$ (observational constraint).
	
	The joint posterior distribution of the NMPs for a given model is the product of the likelihood and the prior distribution of NMPs (Eq. (\ref{eq:bt})). The posterior distribution of individual parameters can be obtained by marginalizing the joint posterior distribution with the remaining model parameters. {If the marginalized posterior distribution of the parameter is narrowed down as compared to the corresponding prior distribution (uniform distribution in this case), then the parameter is said to be constrained by the given data or the likelihood functions. Therefore the narrower distributions of parameters compared to their prior distributions indicate the importance of the likelihood function.} The likelihood function imposes additional constraints on the multi-variate nuclear matter parameters of our model driven by the data. The corner plots for the nuclear matter parameters, which yield the EoSs consistent with the  minimal constraints, are shown in Fig.~\ref{fig1}. The median values of the nuclear matter parameters and the $68\%$ confidence intervals are given in the diagonal plots of the figure. The 68\% confidence intervals for $L_0$, $Q_0$, and $K_{\rm sym,0}$ are significantly smaller than their prior ranges implying these parameters are well constrained by the  low-density EoS for the pure neutron matter. The values of  $J_0$,  and $Q_{\rm sym,0}$are also constrained to some extent.  Except for $L_0-K_{\rm sym,0}$ (r = 0.8) and $L_0-J_0$ (r = 0.65), all other pairs of nuclear matter parameters do not show any visible correlations.
	
	\subsection{Neutron star properties }\label{ns}
	
	The EoSs for $\beta$-equilibrated charge neutral matter in the density ranges $0.5\rho_0$ to the transition density $\rho_{\rm tr}$  are obtained using Taylor expansion with NMPs corresponding to the  posterior distribution as displayed in the Fig.~\ref{fig1}. The calculations are  performed assuming different values for the $\rho_{\rm tr}$ in the range of $1.5 - 2\rho_0$. Each of the EoSs beyond $\rho_{\rm tr}$ is smoothly joined by  a diverse set of EoSs, which are obtained simply by  imposing the  causality condition on the speed of sound by following the Eqs.(\ref{eq-vs}-\ref{eq-preE}).
	The EoS for the density ranges $\rho<0.5\rho_0$ comprises outer and inner crusts. We have. for used the EoS for the outer crust  by Baym-Pethick-Sutherland \cite{Baym:1971pw} in the density range $3.9 \times 10^{-11}\rho_0<\rho<0.0016\rho_0$. We have assumed a polytropic form of the EoS for the inner crust  as follows\cite{Carriere:2002bx}, 
	\be
	p(\varepsilon)= \alpha + \beta \varepsilon^{\frac{4}{3}}. \label{eq-ic}
	\ee
	Here the parameters $\alpha$ and $\beta$ are determined in such a way that the EoS for the inner crust matches with the outer crust at one end and with the outer core at the other end. There is a greater sensitivity to the treatment of crust EoS for neutron stars with mass $\sim 1 M_\odot$~\cite{Fortin:2016hny}. The treatment of crust EoS employed in the present work may introduce the uncertainties of about 50-100 meters in radii of NSs having a mass 1.4$M_\odot$. In Ref. \cite{Piekarewicz2019}, it is shown that the choice of EoS for the inner crust affects both the Love number $k_{2}$ and compactness parameter in such a way that the values of the tidal deformability parameter remain practically unaltered. Once the EoSs for the core and crust are determined, the values of neutron star mass, radius, and tidal deformability corresponding to a given central pressure can be obtained by solving Tolman-Oppenheimer-Volkoff equations~\cite{Oppenheimer:1939ne, Tolman:1939jz}.
	
	\begin{figure}[t]
		\centering
		\includegraphics[width=0.5\textwidth]{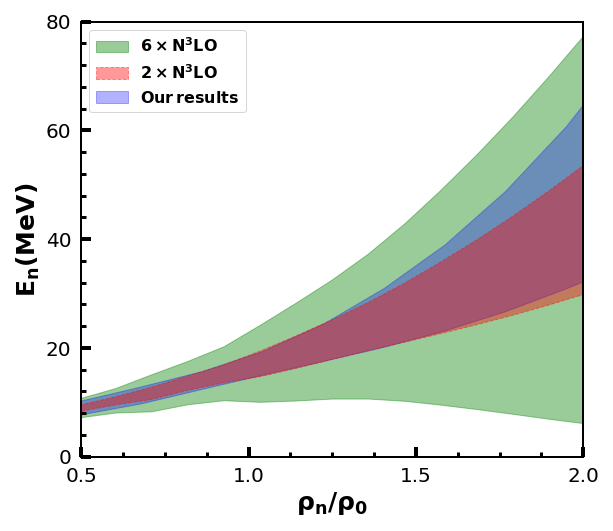}
		\caption{(Color Online) The energy per neutron ($E_n(\rho) = E(\rho,1)$) for pure neutron matter as a function of neutron density. The colored  bands   correspond to 6 $\times$ N$^3$LO (light green), 2 $\times$ N$^3$LO (light red), and 90\% confidence interval for the EoSs (light blue) obtained in our calculation (see text for details).}\label{fig2}
	\end{figure}

	\begin{figure}[htp]
		\centering
		\includegraphics[width=0.5\textwidth]{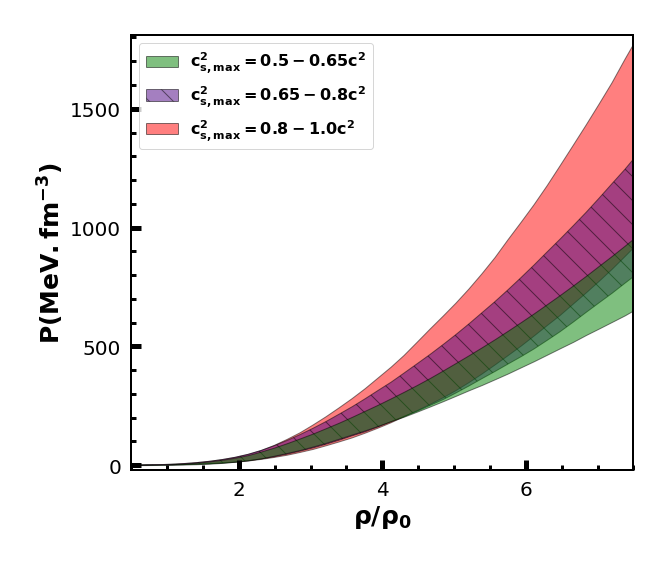}
		\caption{(Color Online) The pressure for $\beta$-equilibrated charge neutral matter as a function of nucleon density. The colored  bands  correspond to 90\% confidence intervals for the EoSs with different ranges of the square of the speed of sound at the center of NS with maximum mass ( $c_{s,\rm max}^2$)(see text for details).}\label{fig3}
	\end{figure}
	
	{The 90\% confidence interval for pure neutron matter EoSs (light blue) for the density range ( $0.5-2.0\rho_0$) is shown in Fig.~\ref{fig2}. For comparison, colored bands correspond to 6 $\times$ N$^3$LO (light green), 2 $\times$ N$^3$LO (light red) from chiral effective field theory are displayed. Our EoSs lie almost in the middle of the 6 $\times$ N$^3$LO band and significantly satisfy the 2 $\times$ N$^3$LO band. This indicates that our EoSs are well fitted with the pure neutron matter EoS from a precise next-to-next-to-leading-order calculation in chiral effective field theory.
		The 90$\%$ confidence interval for the pressure of the $\beta$-equilibrated matter  is plotted as a function of density in Fig.~\ref{fig3}. The results are divided into three groups depending upon the square of sound speed ($c_{s,\rm max}^2$) at the center of NS of its maximum stable mass configuration. The three groups of EoSs  correspond to $c_{s,\rm max}^2=0.5-0.65c^2$, $0.65-0.8c^2$, and $0.8-1.0c^2$ are depicted by different colors. All the EoSs in each group are plotted upto the same density 7$\rho_0$. The overall stiffness of the EoS increases with the $c_{s,\rm max}^2$.}

	\begin{figure*}[htp]
		\centering
		\includegraphics[width=\textwidth]{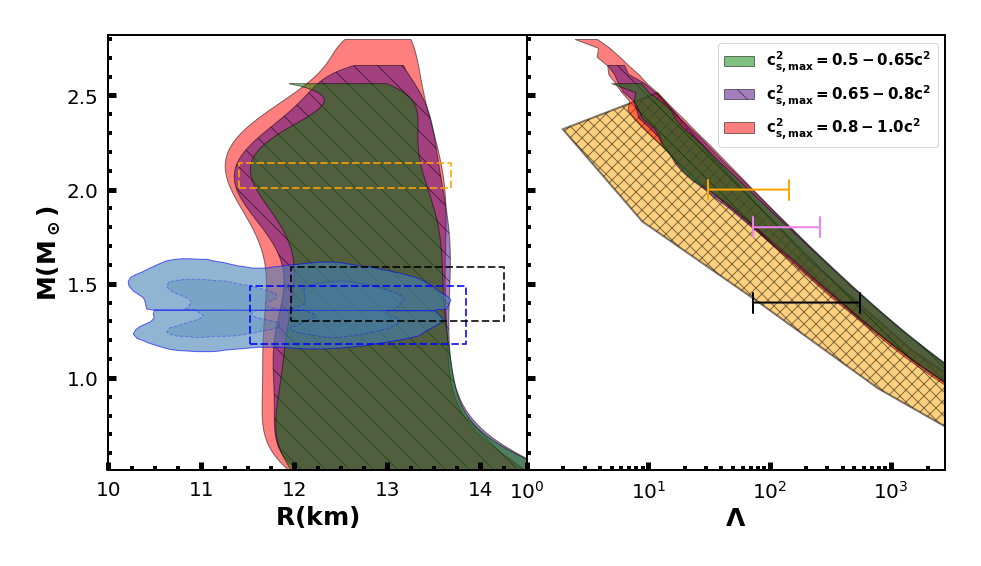}
		\caption{(Color Online) The mass-radius relationship  (left panel)  obtained  for  the EoSs as shown  in Fig.~\ref{fig3}. The outer  and inner gray shaded regions indicate the 90\% (solid) and 50\% (dashed) confidence intervals from the LIGO-Virgo analysis for BNS components of GW170817 event~\cite{GW170817_MR_PEsample, Abbott18a, GWOSC_softx}. The rectangular regions enclosed by  blue \& black  dashed  lines indicate the constraints from the millisecond pulsar PSR J0030+0451 from NICER X-ray observation \cite{Riley2019, Miller2019}, and PSR J0740+6620 orange dashed line \cite{Riley:2021pdl}. The  right panel displays  tidal deformability versus NS mass.   The  orange shaded region  is the  observation for $\Lambda$ with 90\% posterior interval from the LIGO/Virgo Collaboration (GW170817 event) \cite{GW170817_MR_PEsample}. The black line correspond to observational bounds on $\Lambda_{1.4} = 190^{+390}_ {-120}$\cite{Abbott18a}. For the comparison, we have shown violet and gold lines corresponding to $\Lambda_{1.8} = 70-270$, and $\Lambda_{2.0} = 30-150$ obtained from a few well-known theoretical models \cite{Piekarewicz2019, Xie2022}}.
		\label{fig4}
	\end{figure*}
	
	The EoSs displayed in Fig.~\ref{fig3} are employed to obtain the mass-radius relationship for static neutron stars as presented in Fig.~\ref{fig4} (see left panel). For the comparison, we also display the constraints obtained from the GW170817 event  and NICER x-ray observation. The maximum mass of a neutron star lies in the range of $2.1-2.7M_\odot$, and the radius for a neutron star with  mass 1.4$M_\odot$ lies in the range 11.8-14 km. Our  mass-radius relationships exclude  smaller  values of radius for a
	given mass, as predicted by the  GW170817 event. This is due to the choice
	of  priors for the  low-order nuclear matter parameters constrained by
	the experimental data on bulk properties of the finite nuclei.    In the right panel of Fig.~\ref{fig4}, we plot the variations of  tidal deformability as a function of mass. We display the constraints obtained from the GW170817 event for comparison. The value of tidal deformability $\Lambda_{1.4} $ is highlighted\cite{Abbott18a}.
	Further, we depict the constraints on $\Lambda$ for NS of mass  1.8$M_\odot$, and 2.0$M_\odot$  within 90\% CI obtained from ten realistic models that can accurately describe the finite nuclei properties and support the $2M_\odot$ neutron star masses\cite{Piekarewicz2019, Xie2022}. The values of tidal deformability obtained with our minimally constrained EoSs have a reasonable overlap with the ones inferred from the GW170817 event.
	
	\begin{figure*}[htp]
		\includegraphics[height = 25 cm, width=19 cm]{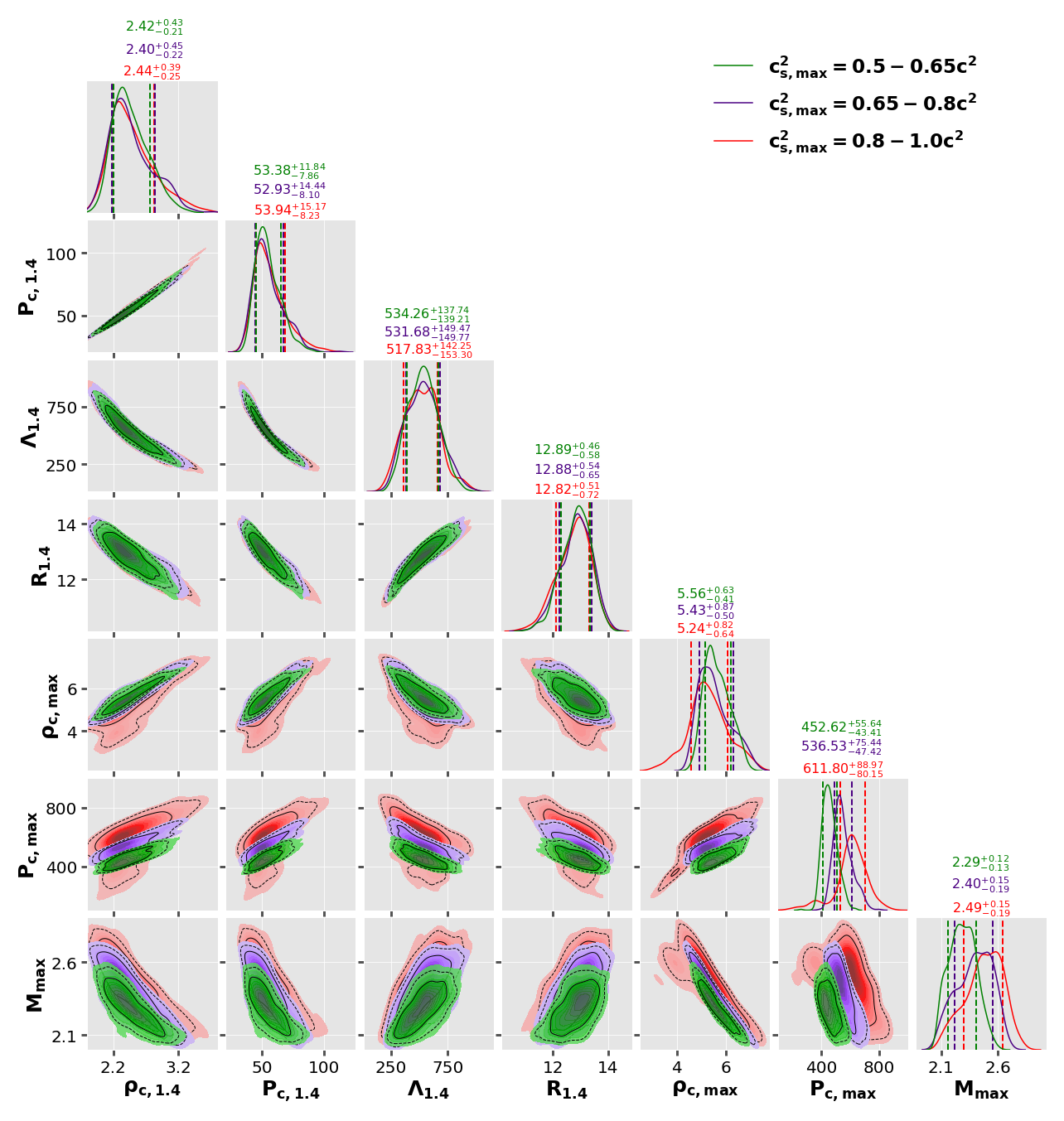}
		\caption{(Color Online) Corner plot for the central density ($\rho_c$ in $\rho_0$) and corresponding pressure ($P_c$ in MeV.fm$^{-3}$) for the neutron star with canonical and maximum mass, radius ($R_{1.4}$ in km),  tidal deformability ($\Lambda_{1.4}$) and maximum mass ($M_{\rm max}$ in $M_\odot$) of NS. The confidence ellipses for two-dimensional posterior distributions are plotted with $1\sigma$(solid line) and $2\sigma$ (dashed line) confidence intervals along the off-diagonal plots. The vertical dashed lines indicate 68\% confidence intervals.} \label{fig5}
	\end{figure*}
	
	In Fig.~\ref{fig5}, we display the corner plot  describing various quantities associated with neutron stars, such as the central density $\rho_c$ and corresponding pressure $P_c$ for the neutron star with canonical ($1.4M_\odot$) and maximum mass, radius $R_{1.4}$,  tidal deformability $\Lambda_{1.4}$  and maximum mass  $M_{\rm max}$. To understand the impact of the high-density EoS, the distributions of all the quantities are segregated into three different groups of stiffness for the EoSs according to the range  of $c_{s,\rm max}^2$ at the higher density part as indicated with different colors. It can be seen from the diagonal plots that the distributions for $\rho_{c,1.4}$, $P_{c,1.4}$, $\Lambda_{1.4}$ and $R_{1.4}$ are more-or-less independent of $c_{s,\rm max}^2$. The median values of $\rho_{c,\rm max}$ decrease with $c_{s,\rm max}^2$, but,  $M_{\rm max}$ and  $P_{c,\rm max}$ increase.
	The $\rho_{c,1.4}$, $P_{c,1.4}$, $R_{1.4}$ and  $\Lambda_{1.4}$ are strongly correlated with each other, the absolute values of Pearson's correlation coefficients being r$\sim 0.86-0.98$. The $\rho_{c,\rm max}$ and $P_{c,\rm max}$ are moderately correlated with the properties of NS with canonical mass ($\mid r \mid \sim  0.70−0.85$), except for $\rho_{c,1.4}$. The  $M_{\rm max}$ show strong correlations with $\rho_{c,\rm max}$ ($\mid r \mid \sim 0.94$), but,  relatively weakly correlated with $P_{c,\rm max} (\mid r \mid\sim 0.78)$. The maximum mass of a stable neutron star seems to be weakly correlated with the properties of NS with canonical mass.
	
	\subsection{ Neutron star properties  and symmetry energy parameters}\label{cor}
	
	\begin{figure*}[t]
		\centering
		\includegraphics[width=\textwidth]{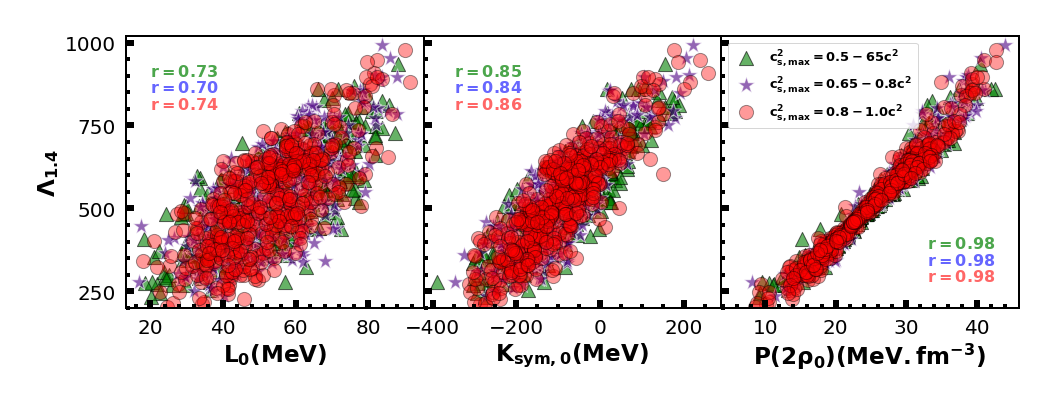}
		\caption{ (Color Online) The tidal deformability ($\Lambda_{1.4}$) as a function of slope parameter ($L_0$) , curvature parameter ($K_{\rm sym,0}$) and the pressure of $\beta$-equilibrated matter (P(2$\rho_0$)).}\label{fig6}
	\end{figure*}
	
	\begin{figure*}[htp]
		\centering
		\includegraphics[width=\textwidth]{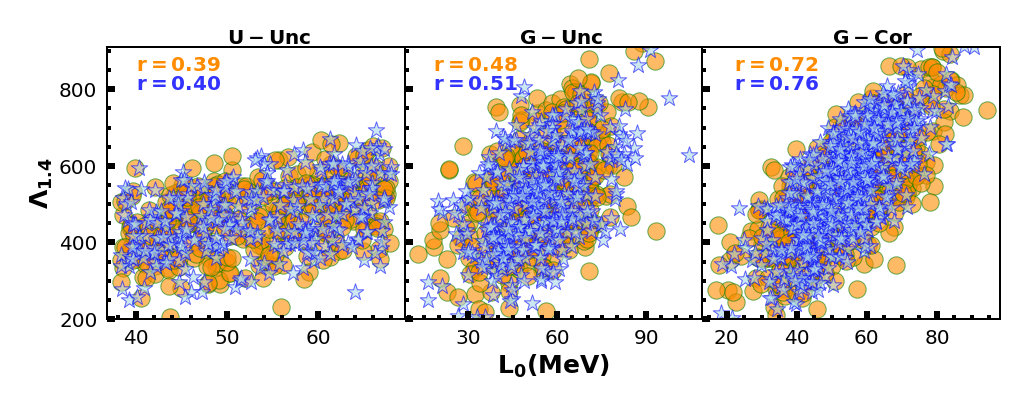}
		\caption{(Color Online) Variations of tidal deformability ($\Lambda_{1.4}$) with slope parameter ($L_0$) for three different distributions of the nuclear matter parameters which are with uniform uncorrelated (U-Unc), Gaussian uncorrelated (G-Unc) and Gaussian correlated posterior distributions (G-Cor) as discussed in the text in details. The results  are obtained by considering the EoSs associated with maximum NS mass $M_{\rm max}$ $\geqslant$ 2.1$M_\odot$. The circle (orange) symbols  represent the results obtained by varying all parameters, whereas the (blue) star symbols represent those obtained by fixing $K_0=240$MeV and $J_0=32$ MeV.}\label{fig7}
	\end{figure*}
	
	The correlations of NS properties with various symmetry energy parameters and the pressure of $\beta-$equilibrated matter have been extensively investigated earlier using non-parametric~\cite{Essick:2021kjb}, parametric~\cite{Tsang:2020lmb, Biswas:2021yge}, and physics-based
	models~\cite{Malik:2020vwo, Reed:2021nqk, Pradhan:2022vdf, Carson:2018xri, Essick:2021kjb} and found to yield the results  which are sometimes at variance, as summarized in  Table III of Ref.~\cite{Kunjipurayil:2022zah}. Some of these studies include constraints imposed by bulk nuclear
	properties, while others have also imposed those through the constraints on nuclear matter parameters assuming them to be independent of each other. We now study the correlations of various NS properties with the $L_0$, $K_{\rm sym,0}$ and $P(2\rho_0)$ using our minimally  constrained EoSs
	and assess  how they  are affected  by the  several factors as listed at
	the beginning of this section.

	We have seen in the previous subsections that our EoSs  yield the various properties of neutron stars within reasonable observational as well as theoretical bounds. As mentioned earlier, the EoSs for $\rho\leqslant\rho_{\rm tr}$ are obtained with the nuclear matter parameters, which are interdependent due to the minimal constraints. We employ these EoSs to  study  the variations  of radius and  tidal deformability with  the slope ($L_0$), curvature ($K_{\rm sym,0}$) and $P(2\rho_0)$  for  NS with mass $1.2 - 1.6M_\odot$. The  parameters $L_0$ and $K_{\rm sym,0}$ determine the density dependence of symmetry energy. We first consider in detail our results for the tidal deformability corresponding to the NS with canonical mass assuming $\rho_{\rm tr} = 2\rho_0$. We show the dependence of tidal deformability $\Lambda_{1.4}$ on the $L_0$, $K_{\rm sym,0}$ and $P(2\rho_0)$ in Fig.~\ref{fig6}. The results for all three groups of EoSs  corresponding to different values of $c_{s,\rm max}^2$ overlap with each other, indicating that the values of $\Lambda_{1.4}$ seem to be  more-or-less insensitive to the behavior of  EoS at high densities $(\rho > \rho_{\rm tr}$). The $\Lambda_{1.4}$ tend to increase with $L_0$, $K_{\rm sym,0}$ and $P(2\rho_0)$.  The values of correlation coefficients, as indicated in Fig.~\ref{fig6}, are practically independent of the choice of $c_{s,\rm max}^2$. Hereafter, we show the results  obtained by combining all three groups of the EoSs.
	
	To study the impact of the interdependence of nuclear matter parameters on the results shown in Fig.~\ref{fig6}, we have generated two different distributions of nuclear matter parameters with the help of their posterior distributions, as shown in Fig.~\ref{fig1}. These distributions of the nuclear matter parameters are (a) uncorrelated Uniform (U-Unc) and (b) uncorrelated Gaussian (G-Unc). The parameters of U-Unc and G-Unc distributions are obtained from the 95\% confidence interval of the marginalized distributions of Fig.~\ref{fig1}. Often, U-Unc and G-Unc distributions have been employed to study the correlations of $\Lambda_{1.4}$ with various symmetry energy parameters \cite{Malik:2020vwo, Pradhan:2022vdf, Carson:2018xri, Tsang:2020lmb}.
	
	In Fig.~\ref{fig7}, we plot the variations of dimensionless tidal deformability $\Lambda_{1.4}$ with $L_0$ for the three different  distributions of nuclear matter parameters as indicated (orange circles). The  extreme right panels are labeled as G-Cor, corresponding to the ones obtained using correlated or joint posterior distribution of nuclear matter parameters. The results shown are obtained by considering the EoSs associated with the maximum NS mass $M_{\rm max}$ $\geqslant$ 2.1$M_\odot$. For the comparison,  the results obtained by fixing the lower order parameters $K_0=240$MeV and $J_0=32$MeV are also displayed (blue stars). In the figure, Pearson's correlation coefficients are given for all three distributions of NMPs.
	The correlations of $\Lambda_{1.4}$ are very sensitive to the choice of the distributions of nuclear matter parameters. The  $\Lambda_{1.4}$ is very weakly correlated with $L_0$ for the case of U-Unc with a correlation coefficient $r\sim 0.39$.. for
	The situation somewhat improves for the case of nuclear matter parameters corresponding to uncorrelated  Gaussian distribution as indicated by G-Unc ($r\sim 0.48$). The posterior distribution (G-Cor) of nuclear matter parameters obtained from minimal constraints (See Fig.~\ref{fig1}) yields relatively stronger correlations of $\Lambda_{1.4}$ with $L_0$ ($r\sim 0.72$).  The correlations  also increase marginally when the values of low-order nuclear matter parameters such as $K_0$ and $J_0$ are kept fixed (blue stars). Evidently, the correlations of $\Lambda_{1.4}$ with $L_0$  depend on the
	various factors, such as the constraints  imposed on the nuclear matter
	parameters that govern the low-density  behavior of the EoSs.  
	
	\begin{figure*}[htp]
		\centering
		\includegraphics[width=\textwidth]{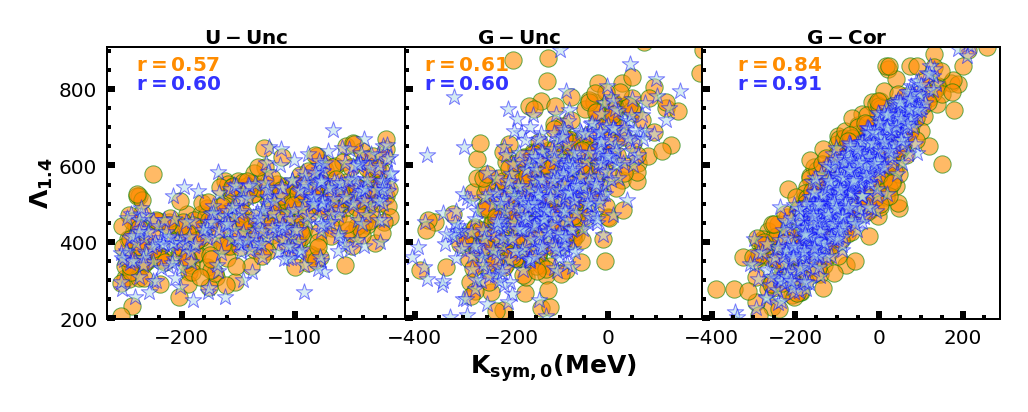}
		\caption{(Color Online)The variation of the tidal deformability ($\Lambda_{1.4}$) with curvature parameter ($K_{\rm sym,0}$) for three different nuclear matter parameter distributions with uniform uncorrelated (U-Unc), Gaussian uncorrelated (G-Unc) and posterior distributions (G-Cor) are discussed in details in the text. The results are shown for those EoSs which are associated with a maximum mass of NS $\geqslant$ 2.1$M_\odot$. The symbols in circles (orange) represent the results obtained by varying all parameters, whereas stars (blue) symbols represent the results obtained when $K_0=240$MeV and $J_0=32$MeV are fixed.}\label{fig8}
	\end{figure*}
	
	\begin{figure*}[htp]
		\centering
		\includegraphics[width=\textwidth]{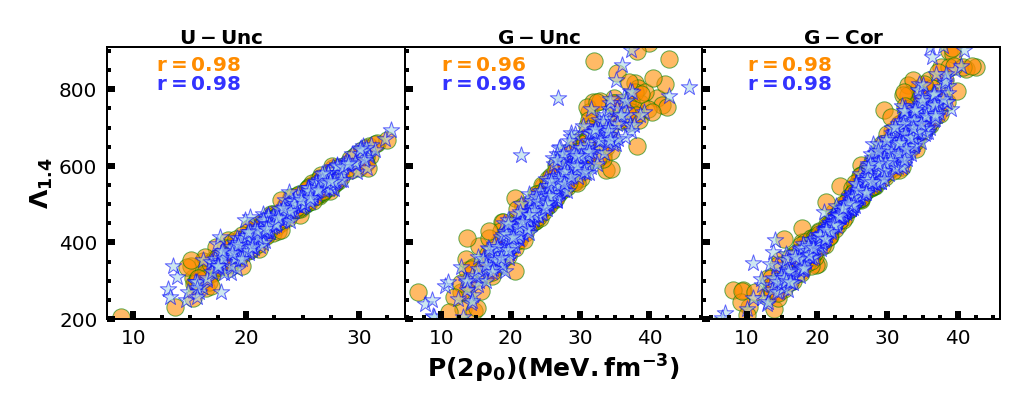}
		\caption{A variation in tidal deformability ($\Lambda_{1.4}$) with the pressure of $\beta$-equilibrated matter at 2$\rho_0$ based on three different nuclear matter parameter distributions, namely uniform uncorrelated (U-Unc), Gaussian uncorrelated (G-Unc) and posterior distributions (G-Cor), are discussed in detail in the text. The results are shown for those EoSs which are associated with a maximum mass of NS $\geqslant$  2.1$M_\odot$. As shown by the circles (orange) symbols, results obtained with all parameters varied, while those obtained with $K_0 = 240$MeV and $J_0 = 32$MeV fixed are represented by the stars (blue) symbols.}\label{fig9}
	\end{figure*}

	The variations of $\Lambda_{1.4}$ with $K_{\rm sym,0}$ for different distributions of nuclear matter parameters are plotted in Fig.~\ref{fig8}. The correlations for $\Lambda_{1.4}$ with $K_{\rm sym,0}$ for different cases are stronger in comparison to those obtained with $L_0$. It appears that the correlations of $\Lambda_{1.4}$ with various symmetry energy parameters are quite sensitive to the distributions of nuclear matter parameters employed.  Our results for the correlations for the U-Unc case are qualitatively similar to those obtained in Refs. \cite{Pradhan:2022vdf, Carson:2018xri} with a similar strategy for nuclear matter parameters but with different models. Similar qualitative trends are observed for the G-Unc \cite{Tsang:2020lmb}. The correlations significantly improve even with the inclusion of minimal constraints, as indicated by G-Cor.  Our results for the case of G-Cor are in harmony with those obtained very recently using about 400 non-relativistic and relativistic mean field models \cite{Carlson:2022nfb}, which demonstrates the impact of low-density EoSs on the properties of NS of canonical mass. It was found that tighter constraints on the  bulk properties of finite nuclei, such as binding energy, charge radii, and isoscalar giant monopole resonance energy, yield stronger  correlations of $\Lambda_{1.4}$ with $L_0$ and $K_{\rm sym,0}$. The correlations of $\Lambda_{1.4}$ with various symmetry energy parameters are stronger only when the nuclear matter parameters evaluated at saturation densities that govern the low-density behavior of EoSs, are appropriately constrained.

	\begin{table*}[htp]
		\caption{\label{tab2} 
			The Pearson's correlation coefficients for  $\Lambda_{1.4}$ with the slope parameter ($L_0$), curvature parameter ($K_{\rm sym,0}$) of the symmetry energy and  the pressure for $\beta$-equilibrated matter at a density $2\rho_0$ ($P(2\rho_0)$) obtained using joint posterior distribution of the nuclear matter parameters. The results are obtained assuming different upper bounds on $\Lambda_{1.4}$ and those are associated with a maximum mass of NS $\geqslant$  2.1$M_\odot$.}
		
		\centering
		%\small\addtolength{\tabcolsep}{0.1pt}
		\begin{ruledtabular}  
			\begin{tabular}{cccc}
				{Upper bound} & $\Lambda_{1.4}-L_0$ & $\Lambda_{1.4}-K_{\rm sym,0}$&$\Lambda_{1.4}-P(2\rho_0)$ \\ 
				on $\Lambda_{1.4}$& &  & \\
				\hline
				400 &0.29 & 0.42&0.88 \\[1.5ex]
				600 & 0.54 & 0.71&0.97  \\[1.5ex]
				800 & 0.67 &0.81 &0.98 \\[1.5ex]
				1000 & 0.72 & 0.84&0.98 \\[1.5ex]
				
			\end{tabular}
		\end{ruledtabular}
	\end{table*}

	\begin{table*}[htp]
		\caption{\label{tab3} 
			The Pearson's correlation coefficients for  $\Lambda_{M}$ and $R_{M}$,  corresponding to different neutron star mass $M$  with the slope parameter ($L_0$), curvature parameter ($K_{\rm sym,0}$) of the symmetry energy and  the pressure for $\beta$-equilibrated matter at a density $2\rho_0$ ($P(2\rho_0)$). The correlation coefficients are obtained  at different transition densities and for the lower bound on  maximum neutron star mass $M_{\rm max}$ $\geqslant$  2.1$M_\odot$.} 
		\centering
		%\small\addtolength{\tabcolsep}{0.1pt}
		\begin{ruledtabular}  
			\begin{tabular}{ccccc|ccc}
				\multirow{2}{*}{$\frac{\rho_{\rm tr}}{\rho_0}$} & \multirow{2}{*}{$\frac{M}{M_\odot}$} &  & {$\Lambda_M$}& & &{$R_M$}&  \\
				\cline{3-8}
				&  & $L_0$ & $K_{\rm sym,0}$ & $P(2\rho_0)$ & $L_0$ & $K_{\rm sym,0}$ & $P(2\rho_0)$ \\
				\cline{3-8}
				\hline
				\multirow{3}{*}{1.5}& 1.2 &0.76 & 0.85&0.91&0.83 & 0.67&0.76 \\[1.5ex]
				&1.4 & 0.70 & 0.81&0.93 &0.79 &0.70 &0.81 \\[1.5ex]
				&1.6 & 0.60 &0.76 &0.91 & 0.72& 0.70&0.86 \\[1.5ex]
				\multirow{3}{*}{1.75}& 1.2 & 0.77 & 0.87&0.94 &0.84 &0.70 &0.81 \\[1.5ex]
				&1.4 &0.71  &0.83 &0.96 &0.80 &0.74 &0.88 \\[1.5ex]
				&1.6 & 0.61 &0.77 &0.97 &0.73 &0.72 & 0.91\\[1.5ex]
				\multirow{3}{*}{2.0}& 1.2 & 0.80 &0.88 &0.97 & 0.85&0.73 &0.86 \\[1.5ex]
				&1.4 & 0.72 &0.84 &0.98 &0.81 &0.75 & 0.91\\[1.5ex]
				&1.6 & 0.65 & 0.80&0.98 & 0.75&0.75 &0.94 \\[1.5ex]
				
			\end{tabular}
		\end{ruledtabular}
	\end{table*}
	
	\begin{table*}[htp]
		\caption{\label{tab4} 
			Same as Table \ref{tab3} but, the results are obtained 
			using the uncertainty on the PNM EoS reduced to 3$\times$ N$^3$L0.}
		\centering
		%\small\addtolength{\tabcolsep}{0.1pt}
		\begin{ruledtabular}  
			\begin{tabular}{ccccc|ccc}
				\multirow{2}{*}{$\frac{\rho_{\rm tr}}{\rho_0}$} & \multirow{2}{*}{$\frac{M}{M_\odot}$} &  & {$\Lambda_M$}& & &{$R_M$}&  \\
				\cline{3-8}
				&  & $L_0$ & $K_{\rm sym,0}$ & $P(2\rho_0)$ & $L_0$ & $K_{\rm sym,0}$ & $P(2\rho_0)$ \\
				\cline{3-8}
				\hline
				\multirow{3}{*}{1.5}& 1.2 &0.47 & 0.73&0.92&0.68 & 0.65&0.74  \\[1.5ex]
				&1.4 & 0.36 & 0.64&0.93 &0.57 &0.66 &0.83 \\[1.5ex]
				&1.6 & 0.26 &0.58 &0.88 & 0.45& 0.65&0.84 \\[1.5ex]
				\multirow{3}{*}{1.75}& 1.2 & 0.50 & 0.76&0.95 &0.70 &0.67 &0.81 \\[1.5ex]
				&1.4 &0.39  &0.72 &0.96 &0.59 &0.68 &0.88 \\[1.5ex]
				&1.6 & 0.31 &0.67 &0.96 &0.49 &0.66 & 0.91\\[1.5ex]
				\multirow{3}{*}{2.0}& 1.2 & 0.51 &0.77 &0.97 & 0.70&0.70 &0.84 \\[1.5ex]
				&1.4 & 0.40 &0.73 &0.98 &0.59 &0.70 & 0.90\\[1.5ex]
				&1.6 & 0.31 & 0.69 & 0.98 & 0.49 & 0.68 & 0.94 \\[1.5ex]
				
			\end{tabular}
		\end{ruledtabular}
	\end{table*}
	
	The variations of $\Lambda_{1.4}$ with the pressure of $\beta$-equilibrated charge neutral matter at twice the saturation densities $P(2\rho_0)$ are plotted in Fig.~\ref{fig9}. The correlations of $\Lambda_{1.4}$ with $P(2\rho_0)$ are quite robust, independent of distributions of nuclear matter parameters. It may be  pointed  out  that the  pressure is related to the density derivative of the energy per particle and would  depend on several
	nuclear matter parameters, except for $\rho=\rho_0$. Its value  at
	$\rho_0$ is  mainly governed by the $L_0$.
	The  strong $\Lambda_{1.4}- P(2\rho_0)$  correlations  may not allow one to reconstruct the EoS of symmetric nuclear matter and the density-dependent symmetry energy very accurately, though they are highly desirable\cite{Imam:2021dbe}. The accurate determination of lower-order nuclear matter parameters from the bulk properties of finite nuclei in conjunction with tighter constraints on $P(2\rho_0)$ may shed some light on the value of high-order nuclear matter parameters.

	The results presented in Figs \ref{fig7} - \ref{fig9} may also be sensitive
	to the range of values for the $\Lambda_{1.4}$ as well as to the  bounds
	on the nuclear matter parameters. In Table \ref{tab2},   we have listed
	the values of Pearson's correlation coefficient for $\Lambda_{1.4} - L_0$,
	$\Lambda_{1.4} - K_{\rm sym,0}$ and $\Lambda_{1.4} - P(2\rho_0)$ obtained
	for the  joint posterior distribution of nuclear matter parameters. The
	correlation coefficients increase with the increase in the upper bound
	on $\Lambda_{1.4}$. In particular, these effects are quite strong for
	the case of $\Lambda_{1.4} - L_0$ and  $\Lambda_{1.4} - K_{\rm sym,0}$.

	So far, we have considered only the case of tidal deformability for  a neutron star with the canonical mass obtained for the  EoSs assuming $\rho_{\rm tr} = 2.0\rho_0$, beyond which the high-density part of the EoS is switched on. We now consider the tidal deformability and radius with neutron star masses M$= 1.2, 1.4$, and $1.6M_\odot$ and study their correlations with symmetry energy slope $L_0$, curvature parameters $K_{\rm sym,0}$ and pressure  $P(2.0\rho_0)$. In Table~\ref{tab3}, we present the values of Pearson's correlation coefficients for $\Lambda_{M}$ and $R_{M}$ with $L_0$, $K_{\rm sym,0}$ and $P(2\rho_0)$. These values of correlation coefficients are calculated  for $\rho_{\rm tr}=1.5, 1.75$ and $2.0\rho_0$ and $M_{\rm max} \geqslant 2.1M_\odot$. The results are obtained for the joint posterior distribution of nuclear matter parameters (Fig.~\ref{fig1}), which determine the EoS at $\rho \leqslant \rho_{\rm tr}$. The correlations increase a little with the $\rho_{\rm tr}=1.5 - 2.0\rho_0$. It is interesting to note that the $\Lambda - L_0$ correlations are moderate, while the $\Lambda - K_{\rm sym,0}$ correlations become relatively stronger which further increase for the $\Lambda -
	P(2\rho_0)$ case. The correlations involving the NS radius display little different trends; the $R - L_0$ correlations are stronger than those for $R- K_{\rm sym,0}$. The correlations of $R-P(2\rho_0)$ are weaker than those for $\Lambda - P(2\rho_0)$. The values of correlation coefficient presented in Table~\ref{tab3} may not be significantly affected by the inclusion of any exotic degrees of freedom beyond 2$\rho_0$, since the EoSs beyond the $\rho_{\rm tr}$ have diverse behavior as they
	are obtained simply by imposing the causality condition.  We have also repeated the calculations to obtain the joint posterior  distribution of nuclear matter parameters by reducing  the uncertainty on the pure neutron matter EoS by a factor of two.  The values of the median and the   68\%  confidence interval  for the  nuclear 
	matter parameters with the reduced uncertainties are $K_0=239_{-32}^{+34}$, $Q_0=-5_{-343}^{+258}$, $J_0=32.19_{-0.9}^{+0.9}$, $L_0=52_{-8}^{+8}$, $K_{\rm sym,0}=-102_{-60}^{+63}$ and $Q_{\rm sym,0}=722_{-510}^{+379}$.
	The values of symmetry energy coefficient $J_0$, slope parameter $L_0$, and curvature parameter $K_{\rm sym,0}$ are now more constrained as compared to those in Fig.~\ref{fig1}.  Similar to Table \ref{tab3}, we have listed the values of Pearson's correlation coefficients for $\Lambda_{M}$ and $R_{M}$ with $L_0$, $K_{\rm sym,0}$ and $P(2\rho_0)$ in Table \ref{tab4}. The values of correlation coefficients show similar trends as listed in Table \ref{tab3}, but the values of correlation coefficients for $\Lambda_{M}$ and $R_{M}$ with $L_0$, $K_{\rm sym,0}$  are reduced. But, the  correlations of the $\Lambda_M$ and $R_M$ with $P(2\rho_0)$
	seem to be more-or-less independent of the  ranges of the nuclear matter parameters,  which once again implies that the NS properties may be more sensitive to the combination of several nuclear matter parameters
	rather than the individual ones.
	{We have also examined the influence of the lower bound on the maximum mass of NS for 2.1 $M_\odot$ to 2.4 $M_\odot$ on our correlation systematics. The correlations do not change significantly; for example, the correlation coefficient value for $\Lambda_{1.4}$ with $L_0$ and $K_{\rm sym,0}$ changes from 0.72 to 0.78 and from 0.84 to 0.86, respectively.}
	
	{We have asset the influence of crust EoS on our correlation systematics. We have repeated our calculations for the correlations of $\Lambda_{1.4}$ and $R_{1.4}$ with $L_0$, $K_{\rm sym,0}$ and $P(2\rho_0)$ for $\rho_{\rm tr}=2\rho_0$ 
		by using different crust EoSs such as NL3$\omega\rho$-L55\cite{Grill:2014aea, Pais:2016xiu} and TM1e\cite{Grill:2014aea, Boukari:2020iut} available in the CompOSE \cite{CompOSE, CompOSECoreTeam:2022ddl}. The results for the correlations involving $\Lambda_{1.4}$ change almost by 1\%. The correlation of $R_{1.4}$ with $L_0$ also remains practically unaffected. However, the correlations of $R_{1.4}$ with $K_{\rm sym,0}$ and $P(2\rho_0)$ improve by 5-10 \%.}

	\section{Conclusions} \label{summary}
	
	We have constructed a large set of minimally constrained EoSs for the NS matter and performed a detailed investigation of the correlations of NS properties with several nuclear matter parameters that determine the density dependence of symmetry energy.
	The joint posterior distribution of nuclear matter parameters that determine the EoSs at low densities ($\rho \leq \rho_{\rm tr}$) is obtained by employing  our minimal constraints within a Bayesian approach. 
	These  EoSs  are consistent with (i) the  pure neutron matter EoS   from a precise next-to-next-to-next-to-leading-order ( N$^{3}$LO) calculation in chiral effective field theory and (ii) empirical ranges of low-order nuclear matter parameters determined by the experimental data on the bulk properties of finite nuclei. 
	The EoSs beyond $\rho_{\rm tr}=1.5-2\rho_0$ are constrained only by imposing the causality condition on the speed of sound. The large set of EoSs so obtained is employed to study the sensitivity of NS properties to the symmetry energy slope parameter $L_0$ and curvature parameter $K_{\rm sym,0}$ as well as to the pressure of $\beta-$equilibrated matter at 2$\rho_0$. The calculations are also performed with uncorrelated Uniform and Gaussian distributions of NMPs that ignore the interdependence among them as present in their joint posterior distribution.
	
	The tidal deformability and radius  of NS, as a function of mass, are evaluated using our minimally constrained EoSs.
	The NS properties at canonical mass  and the maximum NS mass are found to be consistent with the observational constraints. The correlations of tidal deformability and radius for different NS masses ($1.2-1.6 M_\odot$) with the slope and curvature parameter of symmetry energy and the pressure for $\beta-$equilibrated charge neutral matter at $2\rho_0$ are studied.
	We have examined the sensitivity of these correlations 
	to several factors such as, 
	(i) the behavior of the high-density part of the EoS,(ii) the choice of distributions of nuclear matter parameters, their interdependence and  uncertainties,(iii) the lower bound on the maximum mass of the stable neutron stars, (iv)  the value of transition density beyond which the low-density EoSs are smoothly joined with a diverse set of EoSs constrained only by the causality condition on the speed of sound,
	(v) the upper bound on the value of tidal deformability.

	The  tidal deformability is reasonably correlated with the symmetry energy slope parameter $L_0$ for the EoSs obtained from the joint posterior distribution of nuclear matter parameters. The correlation of tidal deformability with $K_{\rm sym,0}$ is slightly stronger. These correlations become even stronger when the priors of lower-order nuclear matter parameters corresponding to incompressibility and symmetry energy coefficients are kept fixed. The correlations become noticeably weaker in the absence of interdependence among nuclear matter parameters. For instance, the Pearson's correlations coefficients for $\Lambda_{1.4}-L_0$ ( $\Lambda_{1.4}-K_{\rm sym,0}$) are $r\sim 0.4(0.6)$ for independent distribution of nuclear matter parameters which become $r\sim0.8(0.9)$ for the joint posterior distribution. This implies that the correlations of NS properties with individual symmetry energy parameters are masked in the absence of appropriate constraints on the EoSs at low densities.  It also partly explains why the outcome of the  similar correlations studied in the earlier publications~\cite{Malik:2020vwo, Pradhan:2022vdf, Carson:2018xri, Tsang:2020lmb, Kunjipurayil:2022zah}  are at variance. These correlations  improve a little bit with the increase in the transition density. The diverse behavior of EoSs at high-density ($\rho \geqslant \rho_{\rm tr}$), as modeled by wide variations of the speed of sound within the causal limit, do not affect the sensitivity of the NS properties to the symmetry energy parameters evaluated at saturation density. 
	The results for the correlation of tidal deformability with the pressure at 2$\rho_0$ are found to be robust, practically independent of all the factors considered.
	The correlations of NS radius with the symmetry energy parameters are also sensitive to various factors considered. The vulnerability of correlations of NS properties with individual parameters of symmetry energy  and, on the  contrary, the robustness of their correlations with $P(2\rho_0)$ needs to be further investigated to pin down the  combination of the optimum number of nuclear matter parameters required to describe the NS properties for the masses $\sim 1.4M_\odot$. With the observations of heavier NS, the correlation between NS properties and symmetry energy parameters presented in this paper may be improved.

	\section{Acknowledgements} 
	The authors are grateful to Sarmistha Banik for  critical reading and the suggestions to improve the presentation of the paper. The authors are thankful to Philippe Landry for providing useful feedback to improve the clarity of the paper. 
	N.K.P. would like to acknowledge the Department of Science and Technology, Ministry of Science and Technology, India, for the support of DST/INSPIRE Fellowship/2019/IF190058. A. M. acknowledges support from the Department of Science and Technology (DST)-Science and Engineering Research Board (SERB) Start-up Research Grant No. SRG/2020/001290. The authors sincerely acknowledge the usage of the analysis software {\tt BILBY} \cite{Ashton2019, Bilby_ref} and open data from GWOSC \cite{GWOSC_softx}. This article has been assigned LIGO Document No. LIGO-P2200392.  {BKA acknowledge
		partial support from the SERB, Department of Science
		and Technology, Government of India with grant no.
		SIR/2022/000566 and CRG/2021/000101, respectively.
		AV would like to acknowledge the CSIR for the support through the CSIR-JRF 09/1026(16303)/2023-EMR-I.}

	%\clearpage
	%\newpage
	%\bibliographystyle{apsrev4-1}
	%\bibliography{library}

\begin{thebibliography}{93}%
	\makeatletter
	\providecommand \@ifxundefined [1]{%
		\@ifx{#1\undefined}
	}%
	\providecommand \@ifnum [1]{%
		\ifnum #1\expandafter \@firstoftwo
		\else \expandafter \@secondoftwo
		\fi
	}%
	\providecommand \@ifx [1]{%
		\ifx #1\expandafter \@firstoftwo
		\else \expandafter \@secondoftwo
		\fi
	}%
	\providecommand \natexlab [1]{#1}%
	\providecommand \enquote  [1]{``#1''}%
	\providecommand \bibnamefont  [1]{#1}%
	\providecommand \bibfnamefont [1]{#1}%
	\providecommand \citenamefont [1]{#1}%
	\providecommand \href@noop [0]{\@secondoftwo}%
	\providecommand \href [0]{\begingroup \@sanitize@url \@href}%
	\providecommand \@href[1]{\@@startlink{#1}\@@href}%
	\providecommand \@@href[1]{\endgroup#1\@@endlink}%
	\providecommand \@sanitize@url [0]{\catcode `\\12\catcode `\$12\catcode
		`\&12\catcode `\#12\catcode `\^12\catcode `\_12\catcode `\%12\relax}%
	\providecommand \@@startlink[1]{}%
	\providecommand \@@endlink[0]{}%
	\providecommand \url  [0]{\begingroup\@sanitize@url \@url }%
	\providecommand \@url [1]{\endgroup\@href {#1}{\urlprefix }}%
	\providecommand \urlprefix  [0]{URL }%
	\providecommand \Eprint [0]{\href }%
	\providecommand \doibase [0]{http://dx.doi.org/}%
	\providecommand \selectlanguage [0]{\@gobble}%
	\providecommand \bibinfo  [0]{\@secondoftwo}%
	\providecommand \bibfield  [0]{\@secondoftwo}%
	\providecommand \translation [1]{[#1]}%
	\providecommand \BibitemOpen [0]{}%
	\providecommand \bibitemStop [0]{}%
	\providecommand \bibitemNoStop [0]{.\EOS\space}%
	\providecommand \EOS [0]{\spacefactor3000\relax}%
	\providecommand \BibitemShut  [1]{\csname bibitem#1\endcsname}%
	\let\auto@bib@innerbib\@empty
	%</preamble>
	\bibitem [{\citenamefont {Oppenheimer}\ and\ \citenamefont
		{Volkoff}(1939)}]{Oppenheimer:1939ne}%
	\BibitemOpen
	\bibfield  {author} {\bibinfo {author} {\bibfnamefont {J.~R.}\ \bibnamefont
			{Oppenheimer}}\ and\ \bibinfo {author} {\bibfnamefont {G.~M.}\ \bibnamefont
			{Volkoff}},\ }\href@noop {} {\bibfield  {journal} {\bibinfo  {journal} {Phys.
				Rev.}\ }\textbf {\bibinfo {volume} {55}},\ \bibinfo {pages} {374} (\bibinfo
		{year} {1939})}\BibitemShut {NoStop}%
	\bibitem [{\citenamefont {Tolman}(1939)}]{Tolman:1939jz}%
	\BibitemOpen
	\bibfield  {author} {\bibinfo {author} {\bibfnamefont {R.~C.}\ \bibnamefont
			{Tolman}},\ }\href@noop {} {\bibfield  {journal} {\bibinfo  {journal} {Phys.
				Rev.}\ }\textbf {\bibinfo {volume} {55}},\ \bibinfo {pages} {364} (\bibinfo
		{year} {1939})}\BibitemShut {NoStop}%
	\bibitem [{\citenamefont {Glendenning}(1992)}]{Glendenning:1992dr}%
	\BibitemOpen
	\bibfield  {author} {\bibinfo {author} {\bibfnamefont {N.~K.}\ \bibnamefont
			{Glendenning}},\ }\href {\doibase 10.1103/PhysRevD.46.4161} {\bibfield
		{journal} {\bibinfo  {journal} {Phys. Rev. D}\ }\textbf {\bibinfo {volume}
			{46}},\ \bibinfo {pages} {4161} (\bibinfo {year} {1992})}\BibitemShut
	{NoStop}%
	\bibitem [{\citenamefont {Huth}\ \emph {et~al.}(2022)\citenamefont {Huth} \emph
		{et~al.}}]{Huth:2021bsp}%
	\BibitemOpen
	\bibfield  {author} {\bibinfo {author} {\bibfnamefont {S.}~\bibnamefont
			{Huth}} \emph {et~al.},\ }\href {\doibase 10.1038/s41586-022-04750-w}
	{\bibfield  {journal} {\bibinfo  {journal} {Nature}\ }\textbf {\bibinfo
			{volume} {606}},\ \bibinfo {pages} {276} (\bibinfo {year}
		{2022})}\BibitemShut {NoStop}%
	\bibitem [{\citenamefont {Aasi}\ \emph {et~al.}(2015)\citenamefont {Aasi} \emph
		{et~al.}}]{LIGOScientific:2014pky}%
	\BibitemOpen
	\bibfield  {author} {\bibinfo {author} {\bibfnamefont {J.}~\bibnamefont
			{Aasi}} \emph {et~al.} (\bibinfo {collaboration} {LIGO Scientific}),\ }\href
	{\doibase 10.1088/0264-9381/32/7/074001} {\bibfield  {journal} {\bibinfo
			{journal} {Class. Quant. Grav.}\ }\textbf {\bibinfo {volume} {32}},\ \bibinfo
		{pages} {074001} (\bibinfo {year} {2015})}\BibitemShut {NoStop}%
	\bibitem [{\citenamefont {Acernese}\ \emph {et~al.}(2015)\citenamefont
		{Acernese} \emph {et~al.}}]{VIRGO:2014yos}%
	\BibitemOpen
	\bibfield  {author} {\bibinfo {author} {\bibfnamefont {F.}~\bibnamefont
			{Acernese}} \emph {et~al.} (\bibinfo {collaboration} {VIRGO}),\ }\href
	{\doibase 10.1088/0264-9381/32/2/024001} {\bibfield  {journal} {\bibinfo
			{journal} {Class. Quant. Grav.}\ }\textbf {\bibinfo {volume} {32}},\ \bibinfo
		{pages} {024001} (\bibinfo {year} {2015})}\BibitemShut {NoStop}%
	\bibitem [{\citenamefont {Abbott~{\sl et
				al.}}(2019{\natexlab{a}})}]{Abbott18a}%
	\BibitemOpen
	\bibfield  {author} {\bibinfo {author} {\bibfnamefont {B.~P.}\ \bibnamefont
			{Abbott~{\sl et al.}}},\ }\href@noop {} {\bibfield  {journal} {\bibinfo
			{journal} {Phys. Rev. X}\ }\textbf {\bibinfo {volume} {9}},\ \bibinfo {pages}
		{011001} (\bibinfo {year} {2019}{\natexlab{a}})}\BibitemShut {NoStop}%
	\bibitem [{\citenamefont {Abbott~{\sl et
				al.}}(2019{\natexlab{b}})}]{Abbott2019}%
	\BibitemOpen
	\bibfield  {author} {\bibinfo {author} {\bibfnamefont {B.~P.}\ \bibnamefont
			{Abbott~{\sl et al.}}},\ }\href@noop {} {\bibfield  {journal} {\bibinfo
			{journal} {Astrophys. J.}\ }\textbf {\bibinfo {volume} {882}},\ \bibinfo
		{pages} {L24} (\bibinfo {year} {2019}{\natexlab{b}})}\BibitemShut {NoStop}%
	\bibitem [{\citenamefont {Abbott~{\sl et al.}}(2020)}]{Abbott2020}%
	\BibitemOpen
	\bibfield  {author} {\bibinfo {author} {\bibfnamefont {B.~P.}\ \bibnamefont
			{Abbott~{\sl et al.}}},\ }\href@noop {} {\bibfield  {journal} {\bibinfo
			{journal} {Astrophys. J.}\ }\textbf {\bibinfo {volume} {892}},\ \bibinfo
		{pages} {L3} (\bibinfo {year} {2020})}\BibitemShut {NoStop}%
	\bibitem [{\citenamefont {Punturo}\ \emph {et~al.}(2010)\citenamefont {Punturo}
		\emph {et~al.}}]{Punturo:2010zz}%
	\BibitemOpen
	\bibfield  {author} {\bibinfo {author} {\bibfnamefont {M.}~\bibnamefont
			{Punturo}} \emph {et~al.},\ }\href {\doibase 10.1088/0264-9381/27/19/194002}
	{\bibfield  {journal} {\bibinfo  {journal} {Class. Quant. Grav.}\ }\textbf
		{\bibinfo {volume} {27}},\ \bibinfo {pages} {194002} (\bibinfo {year}
		{2010})}\BibitemShut {NoStop}%
	\bibitem [{\citenamefont {Reitze}\ \emph {et~al.}(2019)\citenamefont {Reitze}
		\emph {et~al.}}]{Reitze:2019iox}%
	\BibitemOpen
	\bibfield  {author} {\bibinfo {author} {\bibfnamefont {D.}~\bibnamefont
			{Reitze}} \emph {et~al.},\ }\href@noop {} {\bibfield  {journal} {\bibinfo
			{journal} {Bull. Am. Astron. Soc.}\ }\textbf {\bibinfo {volume} {51}},\
		\bibinfo {pages} {035} (\bibinfo {year} {2019})}\BibitemShut {NoStop}%
	\bibitem [{\citenamefont {Abbott~{\sl et al.}}(2017)}]{GW170817}%
	\BibitemOpen
	\bibfield  {author} {\bibinfo {author} {\bibfnamefont {B.~P.}\ \bibnamefont
			{Abbott~{\sl et al.}}},\ }\href@noop {} {\bibfield  {journal} {\bibinfo
			{journal} {Phys. Rev. Lett.}\ }\textbf {\bibinfo {volume} {119}},\ \bibinfo
		{pages} {161101} (\bibinfo {year} {2017})}\BibitemShut {NoStop}%
	\bibitem [{\citenamefont {Malik}\ \emph {et~al.}(2018)\citenamefont {Malik},
		\citenamefont {Alam}, \citenamefont {Fortin}, \citenamefont
		{Provid{\^{e}}ncia}, \citenamefont {Agrawal}, \citenamefont {Jha},
		\citenamefont {Kumar},\ and\ \citenamefont {Patra}}]{Malik2018}%
	\BibitemOpen
	\bibfield  {author} {\bibinfo {author} {\bibfnamefont {T.}~\bibnamefont
			{Malik}}, \bibinfo {author} {\bibfnamefont {N.}~\bibnamefont {Alam}},
		\bibinfo {author} {\bibfnamefont {M.}~\bibnamefont {Fortin}}, \bibinfo
		{author} {\bibfnamefont {C.}~\bibnamefont {Provid{\^{e}}ncia}}, \bibinfo
		{author} {\bibfnamefont {B.~K.}\ \bibnamefont {Agrawal}}, \bibinfo {author}
		{\bibfnamefont {T.~K.}\ \bibnamefont {Jha}}, \bibinfo {author} {\bibfnamefont
			{B.}~\bibnamefont {Kumar}}, \ and\ \bibinfo {author} {\bibfnamefont {S.~K.}\
			\bibnamefont {Patra}},\ }\href@noop {} {\bibfield  {journal} {\bibinfo
			{journal} {Phys. Rev. C}\ }\textbf {\bibinfo {volume} {98}},\ \bibinfo
		{pages} {035804} (\bibinfo {year} {2018})}\BibitemShut {NoStop}%
	\bibitem [{\citenamefont {De}\ \emph {et~al.}(2018)\citenamefont {De},
		\citenamefont {Finstad}, \citenamefont {Lattimer}, \citenamefont {Brown},
		\citenamefont {Berger},\ and\ \citenamefont {Biwer}}]{De18}%
	\BibitemOpen
	\bibfield  {author} {\bibinfo {author} {\bibfnamefont {S.}~\bibnamefont
			{De}}, \bibinfo {author} {\bibfnamefont {D.}~\bibnamefont {Finstad}},
		\bibinfo {author} {\bibfnamefont {J.~M.}\ \bibnamefont {Lattimer}}, \bibinfo
		{author} {\bibfnamefont {D.~A.}\ \bibnamefont {Brown}}, \bibinfo {author}
		{\bibfnamefont {E.}~\bibnamefont {Berger}}, \ and\ \bibinfo {author}
		{\bibfnamefont {C.~M.}\ \bibnamefont {Biwer}},\ }\href@noop {} {\bibfield
		{journal} {\bibinfo  {journal} {Phys. Rev. Lett.}\ }\textbf {\bibinfo
			{volume} {121}},\ \bibinfo {pages} {091102} (\bibinfo {year}
		{2018})}\BibitemShut {NoStop}%
	\bibitem [{\citenamefont {Fattoyev}\ \emph {et~al.}(2018)\citenamefont
		{Fattoyev}, \citenamefont {Piekarewicz},\ and\ \citenamefont
		{Horowitz}}]{Fattoyev2018a}%
	\BibitemOpen
	\bibfield  {author} {\bibinfo {author} {\bibfnamefont {F.~J.}\ \bibnamefont
			{Fattoyev}}, \bibinfo {author} {\bibfnamefont {J.}~\bibnamefont
			{Piekarewicz}}, \ and\ \bibinfo {author} {\bibfnamefont {C.~J.}\ \bibnamefont
			{Horowitz}},\ }\href@noop {} {\bibfield  {journal} {\bibinfo  {journal}
			{Phys. Rev. Lett.}\ }\textbf {\bibinfo {volume} {120}},\ \bibinfo {pages}
		{172702} (\bibinfo {year} {2018})}\BibitemShut {NoStop}%
	\bibitem [{\citenamefont {{Forbes}}\ \emph {et~al.}(2019)\citenamefont
		{{Forbes}}, \citenamefont {{Bose}}, \citenamefont {{Reddy}}, \citenamefont
		{{Zhou}}, \citenamefont {{Mukherjee}},\ and\ \citenamefont
		{{De}}}]{Forbes_etal_withAM2019}%
	\BibitemOpen
	\bibfield  {author} {\bibinfo {author} {\bibfnamefont {M.~M.}\ \bibnamefont
			{{Forbes}}}, \bibinfo {author} {\bibfnamefont {S.}~\bibnamefont {{Bose}}},
		\bibinfo {author} {\bibfnamefont {S.}~\bibnamefont {{Reddy}}}, \bibinfo
		{author} {\bibfnamefont {D.}~\bibnamefont {{Zhou}}}, \bibinfo {author}
		{\bibfnamefont {A.}~\bibnamefont {{Mukherjee}}}, \ and\ \bibinfo {author}
		{\bibfnamefont {S.}~\bibnamefont {{De}}},\ }\href {\doibase
		10.1103/PhysRevD.100.083010} {\bibfield  {journal} {\bibinfo  {journal}
			{\prd}\ }\textbf {\bibinfo {volume} {100}},\ \bibinfo {eid} {083010}
		(\bibinfo {year} {2019})}\BibitemShut {NoStop}%
	\bibitem [{\citenamefont {Landry}\ and\ \citenamefont
		{Essick}(2019)}]{Landry2019}%
	\BibitemOpen
	\bibfield  {author} {\bibinfo {author} {\bibfnamefont {P.}~\bibnamefont
			{Landry}}\ and\ \bibinfo {author} {\bibfnamefont {R.}~\bibnamefont
			{Essick}},\ }\href@noop {} {\bibfield  {journal} {\bibinfo  {journal} {Phys.
				Rev. D}\ }\textbf {\bibinfo {volume} {99}},\ \bibinfo {pages} {084049}
		(\bibinfo {year} {2019})}\BibitemShut {NoStop}%
	\bibitem [{\citenamefont {Piekarewicz}\ and\ \citenamefont
		{Fattoyev}(2019)}]{Piekarewicz2019}%
	\BibitemOpen
	\bibfield  {author} {\bibinfo {author} {\bibfnamefont {J.}~\bibnamefont
			{Piekarewicz}}\ and\ \bibinfo {author} {\bibfnamefont {F.~J.}\ \bibnamefont
			{Fattoyev}},\ }\href@noop {} {\bibfield  {journal} {\bibinfo  {journal}
			{Phys. Rev. C}\ }\textbf {\bibinfo {volume} {99}},\ \bibinfo {pages} {045802}
		(\bibinfo {year} {2019})}\BibitemShut {NoStop}%
	\bibitem [{\citenamefont {Biswas}\ \emph {et~al.}(2021)\citenamefont {Biswas},
		\citenamefont {Char}, \citenamefont {Nandi},\ and\ \citenamefont
		{Bose}}]{Biswas2020}%
	\BibitemOpen
	\bibfield  {author} {\bibinfo {author} {\bibfnamefont {B.}~\bibnamefont
			{Biswas}}, \bibinfo {author} {\bibfnamefont {P.}~\bibnamefont {Char}},
		\bibinfo {author} {\bibfnamefont {R.}~\bibnamefont {Nandi}}, \ and\ \bibinfo
		{author} {\bibfnamefont {S.}~\bibnamefont {Bose}},\ }\href {\doibase
		10.1103/PhysRevD.103.103015} {\bibfield  {journal} {\bibinfo  {journal}
			{Phys. Rev. D}\ }\textbf {\bibinfo {volume} {103}},\ \bibinfo {pages}
		{103015} (\bibinfo {year} {2021})}\BibitemShut {NoStop}%
	\bibitem [{\citenamefont {Thi}\ \emph {et~al.}(2021)\citenamefont {Thi},
		\citenamefont {Mondal},\ and\ \citenamefont {Gulminelli}}]{Thi2021}%
	\BibitemOpen
	\bibfield  {author} {\bibinfo {author} {\bibfnamefont {H.~D.}\ \bibnamefont
			{Thi}}, \bibinfo {author} {\bibfnamefont {C.}~\bibnamefont {Mondal}}, \ and\
		\bibinfo {author} {\bibfnamefont {F.}~\bibnamefont {Gulminelli}},\
	}\href@noop {} {\bibfield  {journal} {\bibinfo  {journal} {Universe}\
		}\textbf {\bibinfo {volume} {7}},\ \bibinfo {pages} {373} (\bibinfo {year}
		{2021})}\BibitemShut {NoStop}%
	\bibitem [{\citenamefont {Miller~{\sl et al.}}(2019)}]{Miller2019}%
	\BibitemOpen
	\bibfield  {author} {\bibinfo {author} {\bibfnamefont {M.~C.}\ \bibnamefont
			{Miller~{\sl et al.}}},\ }\href@noop {} {\bibfield  {journal} {\bibinfo
			{journal} {Astrophys. J.}\ }\textbf {\bibinfo {volume} {887}},\ \bibinfo
		{pages} {L24} (\bibinfo {year} {2019})}\BibitemShut {NoStop}%
	\bibitem [{\citenamefont {Riley~{\sl et al.}}(2019)}]{Riley2019}%
	\BibitemOpen
	\bibfield  {author} {\bibinfo {author} {\bibfnamefont {T.}~\bibnamefont
			{Riley~{\sl et al.}}},\ }\href@noop {} {\bibfield  {journal} {\bibinfo
			{journal} {Astrophys. J. Lett.}\ }\textbf {\bibinfo {volume} {887}},\
		\bibinfo {pages} {L21} (\bibinfo {year} {2019})}\BibitemShut {NoStop}%
	\bibitem [{\citenamefont {Miller~{\sl et al.}}(2021)}]{Miller:2021qha}%
	\BibitemOpen
	\bibfield  {author} {\bibinfo {author} {\bibfnamefont {M.~C.}\ \bibnamefont
			{Miller~{\sl et al.}}},\ }\href@noop {} {\bibfield  {journal} {\bibinfo
			{journal} {Astrophys. J. Lett.}\ }\textbf {\bibinfo {volume} {918}},\
		\bibinfo {pages} {L28} (\bibinfo {year} {2021})}\BibitemShut {NoStop}%
	\bibitem [{\citenamefont {Riley~{\sl et al.}}(2021)}]{Riley:2021pdl}%
	\BibitemOpen
	\bibfield  {author} {\bibinfo {author} {\bibfnamefont {T.~E.}\ \bibnamefont
			{Riley~{\sl et al.}}},\ }\href@noop {} {\bibfield  {journal} {\bibinfo
			{journal} {Astrophys. J. Lett.}\ }\textbf {\bibinfo {volume} {918}},\
		\bibinfo {pages} {L27} (\bibinfo {year} {2021})}\BibitemShut {NoStop}%
	\bibitem [{\citenamefont {Romani}\ \emph {et~al.}(2022)\citenamefont {Romani},
		\citenamefont {Kandel}, \citenamefont {Filippenko}, \citenamefont {Brink},\
		and\ \citenamefont {Zheng}}]{Romani:2022jhd}%
	\BibitemOpen
	\bibfield  {author} {\bibinfo {author} {\bibfnamefont {R.~W.}\ \bibnamefont
			{Romani}}, \bibinfo {author} {\bibfnamefont {D.}~\bibnamefont {Kandel}},
		\bibinfo {author} {\bibfnamefont {A.~V.}\ \bibnamefont {Filippenko}},
		\bibinfo {author} {\bibfnamefont {T.~G.}\ \bibnamefont {Brink}}, \ and\
		\bibinfo {author} {\bibfnamefont {W.}~\bibnamefont {Zheng}},\ }\href
	{\doibase 10.3847/2041-8213/ac8007} {\bibfield  {journal} {\bibinfo
			{journal} {Astrophys. J. Lett.}\ }\textbf {\bibinfo {volume} {934}},\
		\bibinfo {pages} {L17} (\bibinfo {year} {2022})}\BibitemShut {NoStop}%
	\bibitem [{\citenamefont {Linares}\ \emph {et~al.}(2018)\citenamefont
		{Linares}, \citenamefont {Shahbaz},\ and\ \citenamefont
		{Casares}}]{Linares:2018ppq}%
	\BibitemOpen
	\bibfield  {author} {\bibinfo {author} {\bibfnamefont {M.}~\bibnamefont
			{Linares}}, \bibinfo {author} {\bibfnamefont {T.}~\bibnamefont {Shahbaz}}, \
		and\ \bibinfo {author} {\bibfnamefont {J.}~\bibnamefont {Casares}},\ }\href
	{\doibase 10.3847/1538-4357/aabde6} {\bibfield  {journal} {\bibinfo
			{journal} {Astrophys. J.}\ }\textbf {\bibinfo {volume} {859}},\ \bibinfo
		{pages} {54} (\bibinfo {year} {2018})}\BibitemShut {NoStop}%
	\bibitem [{\citenamefont {Hebeler}\ \emph {et~al.}(2013)\citenamefont
		{Hebeler}, \citenamefont {Lattimer}, \citenamefont {Pethick},\ and\
		\citenamefont {Schwenk}}]{Hebeler:2013nza}%
	\BibitemOpen
	\bibfield  {author} {\bibinfo {author} {\bibfnamefont {K.}~\bibnamefont
			{Hebeler}}, \bibinfo {author} {\bibfnamefont {J.~M.}\ \bibnamefont
			{Lattimer}}, \bibinfo {author} {\bibfnamefont {C.~J.}\ \bibnamefont
			{Pethick}}, \ and\ \bibinfo {author} {\bibfnamefont {A.}~\bibnamefont
			{Schwenk}},\ }\href@noop {} {\bibfield  {journal} {\bibinfo  {journal}
			{Astrophys. J.}\ }\textbf {\bibinfo {volume} {773}},\ \bibinfo {pages} {11}
		(\bibinfo {year} {2013})}\BibitemShut {NoStop}%
	\bibitem [{\citenamefont {Li}\ and\ \citenamefont {Steiner}(2006)}]{Li:2005sr}%
	\BibitemOpen
	\bibfield  {author} {\bibinfo {author} {\bibfnamefont {B.-A.}\ \bibnamefont
			{Li}}\ and\ \bibinfo {author} {\bibfnamefont {A.~W.}\ \bibnamefont
			{Steiner}},\ }\href {\doibase 10.1016/j.physletb.2006.09.065} {\bibfield
		{journal} {\bibinfo  {journal} {Phys. Lett. B}\ }\textbf {\bibinfo {volume}
			{642}},\ \bibinfo {pages} {436} (\bibinfo {year} {2006})}\BibitemShut
	{NoStop}%
	\bibitem [{\citenamefont {Lattimer}\ and\ \citenamefont
		{Prakash}(2001)}]{Lattimer:2000nx}%
	\BibitemOpen
	\bibfield  {author} {\bibinfo {author} {\bibfnamefont {J.~M.}\ \bibnamefont
			{Lattimer}}\ and\ \bibinfo {author} {\bibfnamefont {M.}~\bibnamefont
			{Prakash}},\ }\href@noop {} {\bibfield  {journal} {\bibinfo  {journal}
			{Astrophys. J.}\ }\textbf {\bibinfo {volume} {550}},\ \bibinfo {pages} {426}
		(\bibinfo {year} {2001})}\BibitemShut {NoStop}%
	\bibitem [{\citenamefont {Tsang}\ \emph {et~al.}(2019)\citenamefont {Tsang},
		\citenamefont {Tsang}, \citenamefont {Danielewicz}, \citenamefont {Lynch},\
		and\ \citenamefont {Fattoyev}}]{Tsang:2019vxn}%
	\BibitemOpen
	\bibfield  {author} {\bibinfo {author} {\bibfnamefont {C.~Y.}\ \bibnamefont
			{Tsang}}, \bibinfo {author} {\bibfnamefont {M.~B.}\ \bibnamefont {Tsang}},
		\bibinfo {author} {\bibfnamefont {P.}~\bibnamefont {Danielewicz}}, \bibinfo
		{author} {\bibfnamefont {W.~G.}\ \bibnamefont {Lynch}}, \ and\ \bibinfo
		{author} {\bibfnamefont {F.~J.}\ \bibnamefont {Fattoyev}},\ }\href@noop {}
	{\bibfield  {journal} {\bibinfo  {journal} {Phys. Lett. B}\ }\textbf
		{\bibinfo {volume} {796}},\ \bibinfo {pages} {1} (\bibinfo {year}
		{2019})}\BibitemShut {NoStop}%
	\bibitem [{\citenamefont {Tsang}\ \emph {et~al.}(2020)\citenamefont {Tsang},
		\citenamefont {Tsang}, \citenamefont {Danielewicz}, \citenamefont {Lynch},\
		and\ \citenamefont {Fattoyev}}]{Tsang:2020lmb}%
	\BibitemOpen
	\bibfield  {author} {\bibinfo {author} {\bibfnamefont {C.~Y.}\ \bibnamefont
			{Tsang}}, \bibinfo {author} {\bibfnamefont {M.~B.}\ \bibnamefont {Tsang}},
		\bibinfo {author} {\bibfnamefont {P.}~\bibnamefont {Danielewicz}}, \bibinfo
		{author} {\bibfnamefont {W.~G.}\ \bibnamefont {Lynch}}, \ and\ \bibinfo
		{author} {\bibfnamefont {F.~J.}\ \bibnamefont {Fattoyev}},\ }\href@noop {}
	{\bibfield  {journal} {\bibinfo  {journal} {Phys. Rev. C}\ }\textbf {\bibinfo
			{volume} {102}},\ \bibinfo {pages} {045808} (\bibinfo {year}
		{2020})}\BibitemShut {NoStop}%
	\bibitem [{\citenamefont {Patra}\ \emph {et~al.}(2022)\citenamefont {Patra},
		\citenamefont {Imam}, \citenamefont {Agrawal}, \citenamefont {Mukherjee},\
		and\ \citenamefont {Malik}}]{Patra:2022yqc}%
	\BibitemOpen
	\bibfield  {author} {\bibinfo {author} {\bibfnamefont {N.~K.}\ \bibnamefont
			{Patra}}, \bibinfo {author} {\bibfnamefont {S.~M.~A.}\ \bibnamefont {Imam}},
		\bibinfo {author} {\bibfnamefont {B.~K.}\ \bibnamefont {Agrawal}}, \bibinfo
		{author} {\bibfnamefont {A.}~\bibnamefont {Mukherjee}}, \ and\ \bibinfo
		{author} {\bibfnamefont {T.}~\bibnamefont {Malik}},\ }\href {\doibase
		10.1103/PhysRevD.106.043024} {\bibfield  {journal} {\bibinfo  {journal}
			{Phys. Rev. D}\ }\textbf {\bibinfo {volume} {106}},\ \bibinfo {pages}
		{043024} (\bibinfo {year} {2022})}\BibitemShut {NoStop}%
	\bibitem [{\citenamefont {Imam}\ \emph {et~al.}(2022)\citenamefont {Imam},
		\citenamefont {Patra}, \citenamefont {Mondal}, \citenamefont {Malik},\ and\
		\citenamefont {Agrawal}}]{Imam:2021dbe}%
	\BibitemOpen
	\bibfield  {author} {\bibinfo {author} {\bibfnamefont {S.~M.~A.}\
			\bibnamefont {Imam}}, \bibinfo {author} {\bibfnamefont {N.~K.}\ \bibnamefont
			{Patra}}, \bibinfo {author} {\bibfnamefont {C.}~\bibnamefont {Mondal}},
		\bibinfo {author} {\bibfnamefont {T.}~\bibnamefont {Malik}}, \ and\ \bibinfo
		{author} {\bibfnamefont {B.~K.}\ \bibnamefont {Agrawal}},\ }\href@noop {}
	{\bibfield  {journal} {\bibinfo  {journal} {Phys. Rev. C}\ }\textbf {\bibinfo
			{volume} {105}},\ \bibinfo {pages} {015806} (\bibinfo {year}
		{2022})}\BibitemShut {NoStop}%
	\bibitem [{\citenamefont {de~Tovar}\ \emph {et~al.}(2021)\citenamefont
		{de~Tovar}, \citenamefont {Ferreira},\ and\ \citenamefont
		{Provid\^encia}}]{deTovar:2021sjo}%
	\BibitemOpen
	\bibfield  {author} {\bibinfo {author} {\bibfnamefont {P.~B.}\ \bibnamefont
			{de~Tovar}}, \bibinfo {author} {\bibfnamefont {M.}~\bibnamefont {Ferreira}},
		\ and\ \bibinfo {author} {\bibfnamefont {C.}~\bibnamefont {Provid\^encia}},\
	}\href {\doibase 10.1103/PhysRevD.104.123036} {\bibfield  {journal} {\bibinfo
			{journal} {Phys. Rev. D}\ }\textbf {\bibinfo {volume} {104}},\ \bibinfo
		{pages} {123036} (\bibinfo {year} {2021})}\BibitemShut {NoStop}%
	\bibitem [{\citenamefont {Mondal}\ and\ \citenamefont
		{Gulminelli}(2022)}]{Mondal:2021vzt}%
	\BibitemOpen
	\bibfield  {author} {\bibinfo {author} {\bibfnamefont {C.}~\bibnamefont
			{Mondal}}\ and\ \bibinfo {author} {\bibfnamefont {F.}~\bibnamefont
			{Gulminelli}},\ }\href {\doibase 10.1103/PhysRevD.105.083016} {\bibfield
		{journal} {\bibinfo  {journal} {Phys. Rev. D}\ }\textbf {\bibinfo {volume}
			{105}},\ \bibinfo {pages} {083016} (\bibinfo {year} {2022})}\BibitemShut
	{NoStop}%
	\bibitem [{\citenamefont {Alam}\ \emph {et~al.}(2016)\citenamefont {Alam},
		\citenamefont {Agrawal}, \citenamefont {Fortin}, \citenamefont {Pais},
		\citenamefont {Provid\^encia}, \citenamefont {Raduta},\ and\ \citenamefont
		{Sulaksono}}]{Alam:2016cli}%
	\BibitemOpen
	\bibfield  {author} {\bibinfo {author} {\bibfnamefont {N.}~\bibnamefont
			{Alam}}, \bibinfo {author} {\bibfnamefont {B.~K.}\ \bibnamefont {Agrawal}},
		\bibinfo {author} {\bibfnamefont {M.}~\bibnamefont {Fortin}}, \bibinfo
		{author} {\bibfnamefont {H.}~\bibnamefont {Pais}}, \bibinfo {author}
		{\bibfnamefont {C.}~\bibnamefont {Provid\^encia}}, \bibinfo {author}
		{\bibfnamefont {A.~R.}\ \bibnamefont {Raduta}}, \ and\ \bibinfo {author}
		{\bibfnamefont {A.}~\bibnamefont {Sulaksono}},\ }\href {\doibase
		10.1103/PhysRevC.94.052801} {\bibfield  {journal} {\bibinfo  {journal} {Phys.
				Rev. C}\ }\textbf {\bibinfo {volume} {94}},\ \bibinfo {pages} {052801}
		(\bibinfo {year} {2016})}\BibitemShut {NoStop}%
	\bibitem [{\citenamefont {Carson}\ \emph {et~al.}(2019)\citenamefont {Carson},
		\citenamefont {Steiner},\ and\ \citenamefont {Yagi}}]{Carson:2018xri}%
	\BibitemOpen
	\bibfield  {author} {\bibinfo {author} {\bibfnamefont {Z.}~\bibnamefont
			{Carson}}, \bibinfo {author} {\bibfnamefont {A.~W.}\ \bibnamefont {Steiner}},
		\ and\ \bibinfo {author} {\bibfnamefont {K.}~\bibnamefont {Yagi}},\ }\href
	{\doibase 10.1103/PhysRevD.99.043010} {\bibfield  {journal} {\bibinfo
			{journal} {Phys. Rev. D}\ }\textbf {\bibinfo {volume} {99}},\ \bibinfo
		{pages} {043010} (\bibinfo {year} {2019})}\BibitemShut {NoStop}%
	\bibitem [{\citenamefont {G\"uven}\ \emph {et~al.}(2020)\citenamefont
		{G\"uven}, \citenamefont {Bozkurt}, \citenamefont {Khan},\ and\ \citenamefont
		{Margueron}}]{Guven:2020dok}%
	\BibitemOpen
	\bibfield  {author} {\bibinfo {author} {\bibfnamefont {H.}~\bibnamefont
			{G\"uven}}, \bibinfo {author} {\bibfnamefont {K.}~\bibnamefont {Bozkurt}},
		\bibinfo {author} {\bibfnamefont {E.}~\bibnamefont {Khan}}, \ and\ \bibinfo
		{author} {\bibfnamefont {J.}~\bibnamefont {Margueron}},\ }\href {\doibase
		10.1103/PhysRevC.102.015805} {\bibfield  {journal} {\bibinfo  {journal}
			{Phys. Rev. C}\ }\textbf {\bibinfo {volume} {102}},\ \bibinfo {pages}
		{015805} (\bibinfo {year} {2020})}\BibitemShut {NoStop}%
	\bibitem [{\citenamefont {Malik}\ \emph {et~al.}(2020)\citenamefont {Malik},
		\citenamefont {Agrawal}, \citenamefont {Provid\^encia},\ and\ \citenamefont
		{De}}]{Malik:2020vwo}%
	\BibitemOpen
	\bibfield  {author} {\bibinfo {author} {\bibfnamefont {T.}~\bibnamefont
			{Malik}}, \bibinfo {author} {\bibfnamefont {B.~K.}\ \bibnamefont {Agrawal}},
		\bibinfo {author} {\bibfnamefont {C.}~\bibnamefont {Provid\^encia}}, \ and\
		\bibinfo {author} {\bibfnamefont {J.~N.}\ \bibnamefont {De}},\ }\href
	{\doibase 10.1103/PhysRevC.102.052801} {\bibfield  {journal} {\bibinfo
			{journal} {Phys. Rev. C}\ }\textbf {\bibinfo {volume} {102}},\ \bibinfo
		{pages} {052801} (\bibinfo {year} {2020})}\BibitemShut {NoStop}%
	\bibitem [{\citenamefont {Malik}\ and\ \citenamefont
		{Agrawal}(2021)}]{Malik_book}%
	\BibitemOpen
	\bibfield  {author} {\bibinfo {author} {\bibfnamefont {T.}~\bibnamefont
			{Malik}}\ and\ \bibinfo {author} {\bibfnamefont {B.}~\bibnamefont
			{Agrawal}},\ }\href {\doibase ISBN 9780367256104} {\emph {\bibinfo {title} {{
					Constraining the Nuclear Matter EoS from the Properties of Celestial
					Objects}}}},\ Vol.\ \bibinfo {volume} {317}\ (\bibinfo  {publisher} {CRC
		Press},\ \bibinfo {address} {New York, USA},\ \bibinfo {year}
	{2021})\BibitemShut {NoStop}%
	\bibitem [{\citenamefont {Reed}\ \emph {et~al.}(2021)\citenamefont {Reed},
		\citenamefont {Fattoyev}, \citenamefont {Horowitz},\ and\ \citenamefont
		{Piekarewicz}}]{Reed:2021nqk}%
	\BibitemOpen
	\bibfield  {author} {\bibinfo {author} {\bibfnamefont {B.~T.}\ \bibnamefont
			{Reed}}, \bibinfo {author} {\bibfnamefont {F.~J.}\ \bibnamefont {Fattoyev}},
		\bibinfo {author} {\bibfnamefont {C.~J.}\ \bibnamefont {Horowitz}}, \ and\
		\bibinfo {author} {\bibfnamefont {J.}~\bibnamefont {Piekarewicz}},\
	}\href@noop {} {\bibfield  {journal} {\bibinfo  {journal} {Phys. Rev. Lett.}\
		}\textbf {\bibinfo {volume} {126}},\ \bibinfo {pages} {172503} (\bibinfo
		{year} {2021})}\BibitemShut {NoStop}%
	\bibitem [{\citenamefont {Pradhan}\ \emph {et~al.}(2022)\citenamefont
		{Pradhan}, \citenamefont {Chatterjee}, \citenamefont {Lanoye},\ and\
		\citenamefont {Jaikumar}}]{Pradhan:2022vdf}%
	\BibitemOpen
	\bibfield  {author} {\bibinfo {author} {\bibfnamefont {B.~K.}\ \bibnamefont
			{Pradhan}}, \bibinfo {author} {\bibfnamefont {D.}~\bibnamefont {Chatterjee}},
		\bibinfo {author} {\bibfnamefont {M.}~\bibnamefont {Lanoye}}, \ and\ \bibinfo
		{author} {\bibfnamefont {P.}~\bibnamefont {Jaikumar}},\ }\href {\doibase
		10.1103/PhysRevC.106.015805} {\bibfield  {journal} {\bibinfo  {journal}
			{Phys. Rev. C}\ }\textbf {\bibinfo {volume} {106}},\ \bibinfo {pages}
		{015805} (\bibinfo {year} {2022})}\BibitemShut {NoStop}%
	\bibitem [{\citenamefont {Pradhan}\ \emph {et~al.}(2023)\citenamefont
		{Pradhan}, \citenamefont {Chatterjee}, \citenamefont {Gandhi},\ and\
		\citenamefont {Schaffner-Bielich}}]{Pradhan:2022txg}%
	\BibitemOpen
	\bibfield  {author} {\bibinfo {author} {\bibfnamefont {B.~K.}\ \bibnamefont
			{Pradhan}}, \bibinfo {author} {\bibfnamefont {D.}~\bibnamefont {Chatterjee}},
		\bibinfo {author} {\bibfnamefont {R.}~\bibnamefont {Gandhi}}, \ and\ \bibinfo
		{author} {\bibfnamefont {J.}~\bibnamefont {Schaffner-Bielich}},\ }\href
	{\doibase 10.1016/j.nuclphysa.2022.122578} {\bibfield  {journal} {\bibinfo
			{journal} {Nucl. Phys. A}\ }\textbf {\bibinfo {volume} {1030}},\ \bibinfo
		{pages} {122578} (\bibinfo {year} {2023})}\BibitemShut {NoStop}%
	\bibitem [{\citenamefont {Ghosh}\ \emph
		{et~al.}(2022{\natexlab{a}})\citenamefont {Ghosh}, \citenamefont {Pradhan},
		\citenamefont {Chatterjee},\ and\ \citenamefont
		{Schaffner-Bielich}}]{Ghosh:2022lam}%
	\BibitemOpen
	\bibfield  {author} {\bibinfo {author} {\bibfnamefont {S.}~\bibnamefont
			{Ghosh}}, \bibinfo {author} {\bibfnamefont {B.~K.}\ \bibnamefont {Pradhan}},
		\bibinfo {author} {\bibfnamefont {D.}~\bibnamefont {Chatterjee}}, \ and\
		\bibinfo {author} {\bibfnamefont {J.}~\bibnamefont {Schaffner-Bielich}},\
	}\href {\doibase 10.3389/fspas.2022.864294} {\bibfield  {journal} {\bibinfo
			{journal} {Front. Astron. Space Sci.}\ }\textbf {\bibinfo {volume} {9}},\
		\bibinfo {pages} {864294} (\bibinfo {year} {2022}{\natexlab{a}})}\BibitemShut
	{NoStop}%
	\bibitem [{\citenamefont {Ghosh}\ \emph
		{et~al.}(2022{\natexlab{b}})\citenamefont {Ghosh}, \citenamefont
		{Chatterjee},\ and\ \citenamefont {Schaffner-Bielich}}]{Ghosh:2021bvw}%
	\BibitemOpen
	\bibfield  {author} {\bibinfo {author} {\bibfnamefont {S.}~\bibnamefont
			{Ghosh}}, \bibinfo {author} {\bibfnamefont {D.}~\bibnamefont {Chatterjee}}, \
		and\ \bibinfo {author} {\bibfnamefont {J.}~\bibnamefont
			{Schaffner-Bielich}},\ }\href {\doibase 10.1140/epja/s10050-022-00679-w}
	{\bibfield  {journal} {\bibinfo  {journal} {Eur. Phys. J. A}\ }\textbf
		{\bibinfo {volume} {58}},\ \bibinfo {pages} {37} (\bibinfo {year}
		{2022}{\natexlab{b}})}\BibitemShut {NoStop}%
	\bibitem [{\citenamefont {Beznogov}\ and\ \citenamefont
		{Raduta}(2022)}]{Beznogov:2022rri}%
	\BibitemOpen
	\bibfield  {author} {\bibinfo {author} {\bibfnamefont {M.~V.}\ \bibnamefont
			{Beznogov}}\ and\ \bibinfo {author} {\bibfnamefont {A.~R.}\ \bibnamefont
			{Raduta}},\ }\href@noop {} {\  (\bibinfo {year} {2022})},\ \Eprint
	{http://arxiv.org/abs/2212.07168} {arXiv:2212.07168 [nucl-th]} \BibitemShut
	{NoStop}%
	\bibitem [{\citenamefont {Carlson}\ \emph {et~al.}(2023)\citenamefont
		{Carlson}, \citenamefont {Dutra}, \citenamefont {Louren\c{c}o},\ and\
		\citenamefont {Margueron}}]{Carlson:2022nfb}%
	\BibitemOpen
	\bibfield  {author} {\bibinfo {author} {\bibfnamefont {B.~V.}\ \bibnamefont
			{Carlson}}, \bibinfo {author} {\bibfnamefont {M.}~\bibnamefont {Dutra}},
		\bibinfo {author} {\bibfnamefont {O.}~\bibnamefont {Louren\c{c}o}}, \ and\
		\bibinfo {author} {\bibfnamefont {J.}~\bibnamefont {Margueron}},\ }\href
	{\doibase 10.1103/PhysRevC.107.035805} {\bibfield  {journal} {\bibinfo
			{journal} {Phys. Rev. C}\ }\textbf {\bibinfo {volume} {107}},\ \bibinfo
		{pages} {035805} (\bibinfo {year} {2023})}\BibitemShut {NoStop}%
	\bibitem [{\citenamefont {Kunjipurayil}\ \emph {et~al.}(2022)\citenamefont
		{Kunjipurayil}, \citenamefont {Zhao}, \citenamefont {Kumar}, \citenamefont
		{Agrawal},\ and\ \citenamefont {Prakash}}]{Kunjipurayil:2022zah}%
	\BibitemOpen
	\bibfield  {author} {\bibinfo {author} {\bibfnamefont {A.}~\bibnamefont
			{Kunjipurayil}}, \bibinfo {author} {\bibfnamefont {T.}~\bibnamefont {Zhao}},
		\bibinfo {author} {\bibfnamefont {B.}~\bibnamefont {Kumar}}, \bibinfo
		{author} {\bibfnamefont {B.~K.}\ \bibnamefont {Agrawal}}, \ and\ \bibinfo
		{author} {\bibfnamefont {M.}~\bibnamefont {Prakash}},\ }\href {\doibase
		10.1103/PhysRevD.106.063005} {\bibfield  {journal} {\bibinfo  {journal}
			{Phys. Rev. D}\ }\textbf {\bibinfo {volume} {106}},\ \bibinfo {pages}
		{063005} (\bibinfo {year} {2022})}\BibitemShut {NoStop}%
	\bibitem [{\citenamefont {Chatziioannou}\ \emph {et~al.}(2015)\citenamefont
		{Chatziioannou}, \citenamefont {Yagi}, \citenamefont {Klein}, \citenamefont
		{Cornish},\ and\ \citenamefont {Yunes}}]{Chatziioannou:2015uea}%
	\BibitemOpen
	\bibfield  {author} {\bibinfo {author} {\bibfnamefont {K.}~\bibnamefont
			{Chatziioannou}}, \bibinfo {author} {\bibfnamefont {K.}~\bibnamefont {Yagi}},
		\bibinfo {author} {\bibfnamefont {A.}~\bibnamefont {Klein}}, \bibinfo
		{author} {\bibfnamefont {N.}~\bibnamefont {Cornish}}, \ and\ \bibinfo
		{author} {\bibfnamefont {N.}~\bibnamefont {Yunes}},\ }\href {\doibase
		10.1103/PhysRevD.92.104008} {\bibfield  {journal} {\bibinfo  {journal} {Phys.
				Rev. D}\ }\textbf {\bibinfo {volume} {92}},\ \bibinfo {pages} {104008}
		(\bibinfo {year} {2015})}\BibitemShut {NoStop}%
	\bibitem [{\citenamefont {Chatterjee}\ and\ \citenamefont
		{Vida\~na}(2016)}]{Chatterjee:2015pua}%
	\BibitemOpen
	\bibfield  {author} {\bibinfo {author} {\bibfnamefont {D.}~\bibnamefont
			{Chatterjee}}\ and\ \bibinfo {author} {\bibfnamefont {I.}~\bibnamefont
			{Vida\~na}},\ }\href {\doibase 10.1140/epja/i2016-16029-x} {\bibfield
		{journal} {\bibinfo  {journal} {Eur. Phys. J. A}\ }\textbf {\bibinfo {volume}
			{52}},\ \bibinfo {pages} {29} (\bibinfo {year} {2016})}\BibitemShut {NoStop}%
	\bibitem [{\citenamefont {Stone}\ \emph {et~al.}(2021)\citenamefont {Stone},
		\citenamefont {Dexheimer}, \citenamefont {Guichon}, \citenamefont {Thomas},\
		and\ \citenamefont {Typel}}]{Stone:2019blq}%
	\BibitemOpen
	\bibfield  {author} {\bibinfo {author} {\bibfnamefont {J.~R.}\ \bibnamefont
			{Stone}}, \bibinfo {author} {\bibfnamefont {V.}~\bibnamefont {Dexheimer}},
		\bibinfo {author} {\bibfnamefont {P.~A.~M.}\ \bibnamefont {Guichon}},
		\bibinfo {author} {\bibfnamefont {A.~W.}\ \bibnamefont {Thomas}}, \ and\
		\bibinfo {author} {\bibfnamefont {S.}~\bibnamefont {Typel}},\ }\href
	{\doibase 10.1093/mnras/staa4006} {\bibfield  {journal} {\bibinfo  {journal}
			{Mon. Not. Roy. Astron. Soc.}\ }\textbf {\bibinfo {volume} {502}},\ \bibinfo
		{pages} {3476} (\bibinfo {year} {2021})}\BibitemShut {NoStop}%
	\bibitem [{\citenamefont {Chen}\ \emph {et~al.}(2005)\citenamefont {Chen},
		\citenamefont {Ko},\ and\ \citenamefont {Li}}]{Chen:2005ti}%
	\BibitemOpen
	\bibfield  {author} {\bibinfo {author} {\bibfnamefont {L.-W.}\ \bibnamefont
			{Chen}}, \bibinfo {author} {\bibfnamefont {C.~M.}\ \bibnamefont {Ko}}, \ and\
		\bibinfo {author} {\bibfnamefont {B.-A.}\ \bibnamefont {Li}},\ }\href@noop {}
	{\bibfield  {journal} {\bibinfo  {journal} {Phys. Rev. C}\ }\textbf {\bibinfo
			{volume} {72}},\ \bibinfo {pages} {064309} (\bibinfo {year}
		{2005})}\BibitemShut {NoStop}%
	\bibitem [{\citenamefont {Chen}\ \emph {et~al.}(2009)\citenamefont {Chen},
		\citenamefont {Cai}, \citenamefont {Ko}, \citenamefont {Li}, \citenamefont
		{Shen},\ and\ \citenamefont {Xu}}]{Chen:2009wv}%
	\BibitemOpen
	\bibfield  {author} {\bibinfo {author} {\bibfnamefont {L.-W.}\ \bibnamefont
			{Chen}}, \bibinfo {author} {\bibfnamefont {B.-J.}\ \bibnamefont {Cai}},
		\bibinfo {author} {\bibfnamefont {C.~M.}\ \bibnamefont {Ko}}, \bibinfo
		{author} {\bibfnamefont {B.-A.}\ \bibnamefont {Li}}, \bibinfo {author}
		{\bibfnamefont {C.}~\bibnamefont {Shen}}, \ and\ \bibinfo {author}
		{\bibfnamefont {J.}~\bibnamefont {Xu}},\ }\href@noop {} {\bibfield  {journal}
		{\bibinfo  {journal} {Phys. Rev. C}\ }\textbf {\bibinfo {volume} {80}},\
		\bibinfo {pages} {014322} (\bibinfo {year} {2009})}\BibitemShut {NoStop}%
	\bibitem [{\citenamefont {Newton}\ \emph {et~al.}(2014)\citenamefont {Newton},
		\citenamefont {Hooker}, \citenamefont {Gearheart}, \citenamefont {Murphy},
		\citenamefont {Wen}, \citenamefont {Fattoyev},\ and\ \citenamefont
		{Li}}]{Newton:2014iha}%
	\BibitemOpen
	\bibfield  {author} {\bibinfo {author} {\bibfnamefont {W.~G.}\ \bibnamefont
			{Newton}}, \bibinfo {author} {\bibfnamefont {J.}~\bibnamefont {Hooker}},
		\bibinfo {author} {\bibfnamefont {M.}~\bibnamefont {Gearheart}}, \bibinfo
		{author} {\bibfnamefont {K.}~\bibnamefont {Murphy}}, \bibinfo {author}
		{\bibfnamefont {D.-H.}\ \bibnamefont {Wen}}, \bibinfo {author} {\bibfnamefont
			{F.~J.}\ \bibnamefont {Fattoyev}}, \ and\ \bibinfo {author} {\bibfnamefont
			{B.-A.}\ \bibnamefont {Li}},\ }\href@noop {} {\bibfield  {journal} {\bibinfo
			{journal} {Eur. Phys. J. A}\ }\textbf {\bibinfo {volume} {50}},\ \bibinfo
		{pages} {41} (\bibinfo {year} {2014})}\BibitemShut {NoStop}%
	\bibitem [{\citenamefont {Margueron}\ \emph {et~al.}(2018)\citenamefont
		{Margueron}, \citenamefont {{Hoffmann Casali}},\ and\ \citenamefont
		{Gulminelli}}]{Margueron:2017eqc}%
	\BibitemOpen
	\bibfield  {author} {\bibinfo {author} {\bibfnamefont {J.}~\bibnamefont
			{Margueron}}, \bibinfo {author} {\bibfnamefont {R.}~\bibnamefont {{Hoffmann
					Casali}}}, \ and\ \bibinfo {author} {\bibfnamefont {F.}~\bibnamefont
			{Gulminelli}},\ }\href@noop {} {\bibfield  {journal} {\bibinfo  {journal}
			{Phys. Rev. C}\ }\textbf {\bibinfo {volume} {97}},\ \bibinfo {pages} {025805}
		(\bibinfo {year} {2018})}\BibitemShut {NoStop}%
	\bibitem [{\citenamefont {Margueron}\ and\ \citenamefont
		{Gulminelli}(2019)}]{Margueron:2018eob}%
	\BibitemOpen
	\bibfield  {author} {\bibinfo {author} {\bibfnamefont {J.}~\bibnamefont
			{Margueron}}\ and\ \bibinfo {author} {\bibfnamefont {F.}~\bibnamefont
			{Gulminelli}},\ }\href@noop {} {\bibfield  {journal} {\bibinfo  {journal}
			{Phys. Rev. C}\ }\textbf {\bibinfo {volume} {99}},\ \bibinfo {pages} {025806}
		(\bibinfo {year} {2019})}\BibitemShut {NoStop}%
	\bibitem [{\citenamefont {Tews}\ \emph {et~al.}(2018)\citenamefont {Tews},
		\citenamefont {Carlson}, \citenamefont {Gandolfi},\ and\ \citenamefont
		{Reddy}}]{Tews:2018kmu}%
	\BibitemOpen
	\bibfield  {author} {\bibinfo {author} {\bibfnamefont {I.}~\bibnamefont
			{Tews}}, \bibinfo {author} {\bibfnamefont {J.}~\bibnamefont {Carlson}},
		\bibinfo {author} {\bibfnamefont {S.}~\bibnamefont {Gandolfi}}, \ and\
		\bibinfo {author} {\bibfnamefont {S.}~\bibnamefont {Reddy}},\ }\href
	{\doibase 10.3847/1538-4357/aac267} {\bibfield  {journal} {\bibinfo
			{journal} {Astrophys. J.}\ }\textbf {\bibinfo {volume} {860}},\ \bibinfo
		{pages} {149} (\bibinfo {year} {2018})}\BibitemShut {NoStop}%
	\bibitem [{\citenamefont {Gelman~{\sl et al.}}(2013)}]{Gelman2013}%
	\BibitemOpen
	\bibfield  {author} {\bibinfo {author} {\bibfnamefont {A.}~\bibnamefont
			{Gelman~{\sl et al.}}},\ }\href@noop {} {\emph {\bibinfo {title} {{Bayesian
					Data Analysis Third edition}}}}\ (\bibinfo  {publisher} {CRC Press, Boca
		Raton, Florida},\ \bibinfo {year} {2013})\BibitemShut {NoStop}%
	\bibitem [{\citenamefont {Buchner}\ \emph {et~al.}(2014)\citenamefont
		{Buchner}, \citenamefont {Georgakakis}, \citenamefont {Nandra}, \citenamefont
		{Hsu}, \citenamefont {Rangel}, \citenamefont {Brightman}, \citenamefont
		{Merloni}, \citenamefont {Salvato}, \citenamefont {Donley},\ and\
		\citenamefont {Kocevski}}]{Buchner2014}%
	\BibitemOpen
	\bibfield  {author} {\bibinfo {author} {\bibfnamefont {J.}~\bibnamefont
			{Buchner}}, \bibinfo {author} {\bibfnamefont {A.}~\bibnamefont
			{Georgakakis}}, \bibinfo {author} {\bibfnamefont {K.}~\bibnamefont {Nandra}},
		\bibinfo {author} {\bibfnamefont {L.}~\bibnamefont {Hsu}}, \bibinfo {author}
		{\bibfnamefont {C.}~\bibnamefont {Rangel}}, \bibinfo {author} {\bibfnamefont
			{M.}~\bibnamefont {Brightman}}, \bibinfo {author} {\bibfnamefont
			{A.}~\bibnamefont {Merloni}}, \bibinfo {author} {\bibfnamefont
			{M.}~\bibnamefont {Salvato}}, \bibinfo {author} {\bibfnamefont
			{J.}~\bibnamefont {Donley}}, \ and\ \bibinfo {author} {\bibfnamefont
			{D.}~\bibnamefont {Kocevski}},\ }\href {\doibase 10.1051/0004-6361/201322971}
	{\bibfield  {journal} {\bibinfo  {journal} {Astron. Astrophys.}\ }\textbf
		{\bibinfo {volume} {564}},\ \bibinfo {pages} {A125} (\bibinfo {year}
		{2014})}\BibitemShut {NoStop}%
	\bibitem [{\citenamefont {Ashton~{\sl et al.}}(2019)}]{Ashton2019}%
	\BibitemOpen
	\bibfield  {author} {\bibinfo {author} {\bibfnamefont {G.}~\bibnamefont
			{Ashton~{\sl et al.}}},\ }\href@noop {} {\bibfield  {journal} {\bibinfo
			{journal} {Astrophys. J. Suppl. Ser.}\ }\textbf {\bibinfo {volume} {241}},\
		\bibinfo {pages} {27} (\bibinfo {year} {2019})}\BibitemShut {NoStop}%
	\bibitem [{\citenamefont {Chabanat}\ \emph {et~al.}(1998)\citenamefont
		{Chabanat}, \citenamefont {Bonche}, \citenamefont {Haensel}, \citenamefont
		{Meyer},\ and\ \citenamefont {Schaeffer}}]{Chabanat98}%
	\BibitemOpen
	\bibfield  {author} {\bibinfo {author} {\bibfnamefont {E.}~\bibnamefont
			{Chabanat}}, \bibinfo {author} {\bibfnamefont {P.}~\bibnamefont {Bonche}},
		\bibinfo {author} {\bibfnamefont {P.}~\bibnamefont {Haensel}}, \bibinfo
		{author} {\bibfnamefont {J.}~\bibnamefont {Meyer}}, \ and\ \bibinfo {author}
		{\bibfnamefont {R.}~\bibnamefont {Schaeffer}},\ }\href@noop {} {\bibfield
		{journal} {\bibinfo  {journal} {Nucl. Phys. A}\ }\textbf {\bibinfo {volume}
			{635}},\ \bibinfo {pages} {231} (\bibinfo {year} {1998})}\BibitemShut
	{NoStop}%
	\bibitem [{\citenamefont {Chabanat}\ \emph {et~al.}(1997)\citenamefont
		{Chabanat}, \citenamefont {Bonche}, \citenamefont {Haensel}, \citenamefont
		{Meyer},\ and\ \citenamefont {Schaeffer}}]{Chabanat97}%
	\BibitemOpen
	\bibfield  {author} {\bibinfo {author} {\bibfnamefont {E.}~\bibnamefont
			{Chabanat}}, \bibinfo {author} {\bibfnamefont {P.}~\bibnamefont {Bonche}},
		\bibinfo {author} {\bibfnamefont {P.}~\bibnamefont {Haensel}}, \bibinfo
		{author} {\bibfnamefont {J.}~\bibnamefont {Meyer}}, \ and\ \bibinfo {author}
		{\bibfnamefont {R.}~\bibnamefont {Schaeffer}},\ }\href@noop {} {\bibfield
		{journal} {\bibinfo  {journal} {Nucl. Phys. A}\ }\textbf {\bibinfo {volume}
			{627}},\ \bibinfo {pages} {710} (\bibinfo {year} {1997})}\BibitemShut
	{NoStop}%
	\bibitem [{\citenamefont {Malik}\ \emph {et~al.}(2019)\citenamefont {Malik},
		\citenamefont {Agrawal}, \citenamefont {De}, \citenamefont {Samaddar},
		\citenamefont {Provid{\^{e}}ncia}, \citenamefont {Mondal},\ and\
		\citenamefont {Jha}}]{Malik:2019whk}%
	\BibitemOpen
	\bibfield  {author} {\bibinfo {author} {\bibfnamefont {T.}~\bibnamefont
			{Malik}}, \bibinfo {author} {\bibfnamefont {B.~K.}\ \bibnamefont {Agrawal}},
		\bibinfo {author} {\bibfnamefont {J.~N.}\ \bibnamefont {De}}, \bibinfo
		{author} {\bibfnamefont {S.~K.}\ \bibnamefont {Samaddar}}, \bibinfo {author}
		{\bibfnamefont {C.}~\bibnamefont {Provid{\^{e}}ncia}}, \bibinfo {author}
		{\bibfnamefont {C.}~\bibnamefont {Mondal}}, \ and\ \bibinfo {author}
		{\bibfnamefont {T.~K.}\ \bibnamefont {Jha}},\ }\href@noop {} {\bibfield
		{journal} {\bibinfo  {journal} {Phys. Rev. C}\ }\textbf {\bibinfo {volume}
			{99}},\ \bibinfo {pages} {052801} (\bibinfo {year} {2019})}\BibitemShut
	{NoStop}%
	\bibitem [{\citenamefont {Mondal}\ \emph {et~al.}(2015)\citenamefont {Mondal},
		\citenamefont {Agrawal},\ and\ \citenamefont {De}}]{Mondal:2015tfa}%
	\BibitemOpen
	\bibfield  {author} {\bibinfo {author} {\bibfnamefont {C.}~\bibnamefont
			{Mondal}}, \bibinfo {author} {\bibfnamefont {B.~K.}\ \bibnamefont {Agrawal}},
		\ and\ \bibinfo {author} {\bibfnamefont {J.~N.}\ \bibnamefont {De}},\ }\href
	{\doibase 10.1103/PhysRevC.92.024302} {\bibfield  {journal} {\bibinfo
			{journal} {Phys. Rev. C}\ }\textbf {\bibinfo {volume} {92}},\ \bibinfo
		{pages} {024302} (\bibinfo {year} {2015})}\BibitemShut {NoStop}%
	\bibitem [{\citenamefont {Mondal}\ \emph {et~al.}(2016)\citenamefont {Mondal},
		\citenamefont {Agrawal}, \citenamefont {De},\ and\ \citenamefont
		{Samaddar}}]{Mondal:2016roo}%
	\BibitemOpen
	\bibfield  {author} {\bibinfo {author} {\bibfnamefont {C.}~\bibnamefont
			{Mondal}}, \bibinfo {author} {\bibfnamefont {B.~K.}\ \bibnamefont {Agrawal}},
		\bibinfo {author} {\bibfnamefont {J.~N.}\ \bibnamefont {De}}, \ and\ \bibinfo
		{author} {\bibfnamefont {S.~K.}\ \bibnamefont {Samaddar}},\ }\href {\doibase
		10.1103/PhysRevC.93.044328} {\bibfield  {journal} {\bibinfo  {journal} {Phys.
				Rev. C}\ }\textbf {\bibinfo {volume} {93}},\ \bibinfo {pages} {044328}
		(\bibinfo {year} {2016})}\BibitemShut {NoStop}%
	\bibitem [{\citenamefont {Sulaksono}\ \emph {et~al.}(2009)\citenamefont
		{Sulaksono}, \citenamefont {Buervenich}, \citenamefont {Reinhard},\ and\
		\citenamefont {Maruhn}}]{Sulaksono:2009rn}%
	\BibitemOpen
	\bibfield  {author} {\bibinfo {author} {\bibfnamefont {A.}~\bibnamefont
			{Sulaksono}}, \bibinfo {author} {\bibfnamefont {T.~J.}\ \bibnamefont
			{Buervenich}}, \bibinfo {author} {\bibfnamefont {P.~G.}\ \bibnamefont
			{Reinhard}}, \ and\ \bibinfo {author} {\bibfnamefont {J.~A.}\ \bibnamefont
			{Maruhn}},\ }\href {\doibase 10.1103/PhysRevC.79.044306} {\bibfield
		{journal} {\bibinfo  {journal} {Phys. Rev. C}\ }\textbf {\bibinfo {volume}
			{79}},\ \bibinfo {pages} {044306} (\bibinfo {year} {2009})}\BibitemShut
	{NoStop}%
	\bibitem [{\citenamefont {Essick}\ \emph
		{et~al.}(2021{\natexlab{a}})\citenamefont {Essick}, \citenamefont {Landry},
		\citenamefont {Schwenk},\ and\ \citenamefont {Tews}}]{Essick:2021ezp}%
	\BibitemOpen
	\bibfield  {author} {\bibinfo {author} {\bibfnamefont {R.}~\bibnamefont
			{Essick}}, \bibinfo {author} {\bibfnamefont {P.}~\bibnamefont {Landry}},
		\bibinfo {author} {\bibfnamefont {A.}~\bibnamefont {Schwenk}}, \ and\
		\bibinfo {author} {\bibfnamefont {I.}~\bibnamefont {Tews}},\ }\href {\doibase
		10.1103/PhysRevC.104.065804} {\bibfield  {journal} {\bibinfo  {journal}
			{Phys. Rev. C}\ }\textbf {\bibinfo {volume} {104}},\ \bibinfo {pages}
		{065804} (\bibinfo {year} {2021}{\natexlab{a}})}\BibitemShut {NoStop}%
	\bibitem [{\citenamefont {Garg}\ and\ \citenamefont
		{Col\`o}(2018)}]{Garg:2018uam}%
	\BibitemOpen
	\bibfield  {author} {\bibinfo {author} {\bibfnamefont {U.}~\bibnamefont
			{Garg}}\ and\ \bibinfo {author} {\bibfnamefont {G.}~\bibnamefont {Col\`o}},\
	}\href {\doibase 10.1016/j.ppnp.2018.03.001} {\bibfield  {journal} {\bibinfo
			{journal} {Prog. Part. Nucl. Phys.}\ }\textbf {\bibinfo {volume} {101}},\
		\bibinfo {pages} {55} (\bibinfo {year} {2018})}\BibitemShut {NoStop}%
	\bibitem [{\citenamefont {Agrawal}\ \emph {et~al.}(2005)\citenamefont
		{Agrawal}, \citenamefont {Shlomo},\ and\ \citenamefont
		{Au}}]{Agrawal:2005ix}%
	\BibitemOpen
	\bibfield  {author} {\bibinfo {author} {\bibfnamefont {B.~K.}\ \bibnamefont
			{Agrawal}}, \bibinfo {author} {\bibfnamefont {S.}~\bibnamefont {Shlomo}}, \
		and\ \bibinfo {author} {\bibfnamefont {V.~K.}\ \bibnamefont {Au}},\ }\href
	{\doibase 10.1103/PhysRevC.72.014310} {\bibfield  {journal} {\bibinfo
			{journal} {Phys. Rev. C}\ }\textbf {\bibinfo {volume} {72}},\ \bibinfo
		{pages} {014310} (\bibinfo {year} {2005})}\BibitemShut {NoStop}%
	\bibitem [{\citenamefont {Adhikari}\ \emph {et~al.}(2022)\citenamefont
		{Adhikari} \emph {et~al.}}]{CREX:2022kgg}%
	\BibitemOpen
	\bibfield  {author} {\bibinfo {author} {\bibfnamefont {D.}~\bibnamefont
			{Adhikari}} \emph {et~al.} (\bibinfo {collaboration} {CREX}),\ }\href
	{\doibase 10.1103/PhysRevLett.129.042501} {\bibfield  {journal} {\bibinfo
			{journal} {Phys. Rev. Lett.}\ }\textbf {\bibinfo {volume} {129}},\ \bibinfo
		{pages} {042501} (\bibinfo {year} {2022})}\BibitemShut {NoStop}%
	\bibitem [{\citenamefont {Adhikari}\ \emph {et~al.}(2021)\citenamefont
		{Adhikari} \emph {et~al.}}]{PREX:2021umo}%
	\BibitemOpen
	\bibfield  {author} {\bibinfo {author} {\bibfnamefont {D.}~\bibnamefont
			{Adhikari}} \emph {et~al.} (\bibinfo {collaboration} {PREX}),\ }\href
	{\doibase 10.1103/PhysRevLett.126.172502} {\bibfield  {journal} {\bibinfo
			{journal} {Phys. Rev. Lett.}\ }\textbf {\bibinfo {volume} {126}},\ \bibinfo
		{pages} {172502} (\bibinfo {year} {2021})}\BibitemShut {NoStop}%
	\bibitem [{\citenamefont {Reed}\ and\ \citenamefont
		{Horowitz}(2020)}]{Reed:2019ezm}%
	\BibitemOpen
	\bibfield  {author} {\bibinfo {author} {\bibfnamefont {B.}~\bibnamefont
			{Reed}}\ and\ \bibinfo {author} {\bibfnamefont {C.~J.}\ \bibnamefont
			{Horowitz}},\ }\href {\doibase 10.1103/PhysRevC.101.045803} {\bibfield
		{journal} {\bibinfo  {journal} {Phys. Rev. C}\ }\textbf {\bibinfo {volume}
			{101}},\ \bibinfo {pages} {045803} (\bibinfo {year} {2020})}\BibitemShut
	{NoStop}%
	\bibitem [{\citenamefont {Essick}\ \emph
		{et~al.}(2021{\natexlab{b}})\citenamefont {Essick}, \citenamefont {Tews},
		\citenamefont {Landry},\ and\ \citenamefont {Schwenk}}]{Essick:2021kjb}%
	\BibitemOpen
	\bibfield  {author} {\bibinfo {author} {\bibfnamefont {R.}~\bibnamefont
			{Essick}}, \bibinfo {author} {\bibfnamefont {I.}~\bibnamefont {Tews}},
		\bibinfo {author} {\bibfnamefont {P.}~\bibnamefont {Landry}}, \ and\ \bibinfo
		{author} {\bibfnamefont {A.}~\bibnamefont {Schwenk}},\ }\href {\doibase
		10.1103/PhysRevLett.127.192701} {\bibfield  {journal} {\bibinfo  {journal}
			{Phys. Rev. Lett.}\ }\textbf {\bibinfo {volume} {127}},\ \bibinfo {pages}
		{192701} (\bibinfo {year} {2021}{\natexlab{b}})}\BibitemShut {NoStop}%
	\bibitem [{\citenamefont {Reinhard}\ \emph {et~al.}(2021)\citenamefont
		{Reinhard}, \citenamefont {Roca-Maza},\ and\ \citenamefont
		{Nazarewicz}}]{Reinhard:2021utv}%
	\BibitemOpen
	\bibfield  {author} {\bibinfo {author} {\bibfnamefont {P.-G.}\ \bibnamefont
			{Reinhard}}, \bibinfo {author} {\bibfnamefont {X.}~\bibnamefont {Roca-Maza}},
		\ and\ \bibinfo {author} {\bibfnamefont {W.}~\bibnamefont {Nazarewicz}},\
	}\href {\doibase 10.1103/PhysRevLett.127.232501} {\bibfield  {journal}
		{\bibinfo  {journal} {Phys. Rev. Lett.}\ }\textbf {\bibinfo {volume} {127}},\
		\bibinfo {pages} {232501} (\bibinfo {year} {2021})}\BibitemShut {NoStop}%
	\bibitem [{\citenamefont {Tagami}\ \emph {et~al.}(2022)\citenamefont {Tagami},
		\citenamefont {Wakasa},\ and\ \citenamefont {Yahiro}}]{TAGAMI2022}%
	\BibitemOpen
	\bibfield  {author} {\bibinfo {author} {\bibfnamefont {S.}~\bibnamefont
			{Tagami}}, \bibinfo {author} {\bibfnamefont {T.}~\bibnamefont {Wakasa}}, \
		and\ \bibinfo {author} {\bibfnamefont {M.}~\bibnamefont {Yahiro}},\ }\href
	{\doibase https://doi.org/10.1016/j.rinp.2022.106037} {\bibfield  {journal}
		{\bibinfo  {journal} {Results in Physics}\ }\textbf {\bibinfo {volume}
			{43}},\ \bibinfo {pages} {106037} (\bibinfo {year} {2022})}\BibitemShut
	{NoStop}%
	\bibitem [{\citenamefont {Ferreira}\ and\ \citenamefont
		{Provid\^encia}(2021)}]{Ferreira:2021pni}%
	\BibitemOpen
	\bibfield  {author} {\bibinfo {author} {\bibfnamefont {M.}~\bibnamefont
			{Ferreira}}\ and\ \bibinfo {author} {\bibfnamefont {C.}~\bibnamefont
			{Provid\^encia}},\ }\href {\doibase 10.1103/PhysRevD.104.063006} {\bibfield
		{journal} {\bibinfo  {journal} {Phys. Rev. D}\ }\textbf {\bibinfo {volume}
			{104}},\ \bibinfo {pages} {063006} (\bibinfo {year} {2021})}\BibitemShut
	{NoStop}%
	\bibitem [{\citenamefont {Dutra}\ \emph {et~al.}(2012)\citenamefont {Dutra},
		\citenamefont {Lourenco}, \citenamefont {Sa~Martins}, \citenamefont
		{Delfino}, \citenamefont {Stone},\ and\ \citenamefont
		{Stevenson}}]{Dutra:2012mb}%
	\BibitemOpen
	\bibfield  {author} {\bibinfo {author} {\bibfnamefont {M.}~\bibnamefont
			{Dutra}}, \bibinfo {author} {\bibfnamefont {O.}~\bibnamefont {Lourenco}},
		\bibinfo {author} {\bibfnamefont {J.~S.}\ \bibnamefont {Sa~Martins}},
		\bibinfo {author} {\bibfnamefont {A.}~\bibnamefont {Delfino}}, \bibinfo
		{author} {\bibfnamefont {J.~R.}\ \bibnamefont {Stone}}, \ and\ \bibinfo
		{author} {\bibfnamefont {P.~D.}\ \bibnamefont {Stevenson}},\ }\href {\doibase
		10.1103/PhysRevC.85.035201} {\bibfield  {journal} {\bibinfo  {journal} {Phys.
				Rev. C}\ }\textbf {\bibinfo {volume} {85}},\ \bibinfo {pages} {035201}
		(\bibinfo {year} {2012})}\BibitemShut {NoStop}%
	\bibitem [{\citenamefont {Dutra}\ \emph {et~al.}(2014)\citenamefont {Dutra},
		\citenamefont {Louren\c{c}o}, \citenamefont {Avancini}, \citenamefont
		{Carlson}, \citenamefont {Delfino}, \citenamefont {Menezes}, \citenamefont
		{Provid\^encia}, \citenamefont {Typel},\ and\ \citenamefont
		{Stone}}]{Dutra:2014qga}%
	\BibitemOpen
	\bibfield  {author} {\bibinfo {author} {\bibfnamefont {M.}~\bibnamefont
			{Dutra}}, \bibinfo {author} {\bibfnamefont {O.}~\bibnamefont {Louren\c{c}o}},
		\bibinfo {author} {\bibfnamefont {S.~S.}\ \bibnamefont {Avancini}}, \bibinfo
		{author} {\bibfnamefont {B.~V.}\ \bibnamefont {Carlson}}, \bibinfo {author}
		{\bibfnamefont {A.}~\bibnamefont {Delfino}}, \bibinfo {author} {\bibfnamefont
			{D.~P.}\ \bibnamefont {Menezes}}, \bibinfo {author} {\bibfnamefont
			{C.}~\bibnamefont {Provid\^encia}}, \bibinfo {author} {\bibfnamefont
			{S.}~\bibnamefont {Typel}}, \ and\ \bibinfo {author} {\bibfnamefont {J.~R.}\
			\bibnamefont {Stone}},\ }\href {\doibase 10.1103/PhysRevC.90.055203}
	{\bibfield  {journal} {\bibinfo  {journal} {Phys. Rev. C}\ }\textbf {\bibinfo
			{volume} {90}},\ \bibinfo {pages} {055203} (\bibinfo {year}
		{2014})}\BibitemShut {NoStop}%
	\bibitem [{\citenamefont {Mondal}\ \emph {et~al.}(2017)\citenamefont {Mondal},
		\citenamefont {Agrawal}, \citenamefont {De}, \citenamefont {Samaddar},
		\citenamefont {Centelles},\ and\ \citenamefont
		{Vi{\~{n}}as}}]{Mondal:2017hnh}%
	\BibitemOpen
	\bibfield  {author} {\bibinfo {author} {\bibfnamefont {C.}~\bibnamefont
			{Mondal}}, \bibinfo {author} {\bibfnamefont {B.~K.}\ \bibnamefont {Agrawal}},
		\bibinfo {author} {\bibfnamefont {J.~N.}\ \bibnamefont {De}}, \bibinfo
		{author} {\bibfnamefont {S.~K.}\ \bibnamefont {Samaddar}}, \bibinfo {author}
		{\bibfnamefont {M.}~\bibnamefont {Centelles}}, \ and\ \bibinfo {author}
		{\bibfnamefont {X.}~\bibnamefont {Vi{\~{n}}as}},\ }\href@noop {} {\bibfield
		{journal} {\bibinfo  {journal} {Phys. Rev. C}\ }\textbf {\bibinfo {volume}
			{96}},\ \bibinfo {pages} {021302} (\bibinfo {year} {2017})}\BibitemShut
	{NoStop}%
	\bibitem [{\citenamefont {Lattimer}(2021)}]{Lattimer:2021emm}%
	\BibitemOpen
	\bibfield  {author} {\bibinfo {author} {\bibfnamefont {J.~M.}\ \bibnamefont
			{Lattimer}},\ }\href {\doibase 10.1146/annurev-nucl-102419-124827} {\bibfield
		{journal} {\bibinfo  {journal} {Ann. Rev. Nucl. Part. Sci.}\ }\textbf
		{\bibinfo {volume} {71}},\ \bibinfo {pages} {433} (\bibinfo {year}
		{2021})}\BibitemShut {NoStop}%
	\bibitem [{\citenamefont {Baym}\ \emph {et~al.}(1971)\citenamefont {Baym},
		\citenamefont {Pethick},\ and\ \citenamefont {Sutherland}}]{Baym:1971pw}%
	\BibitemOpen
	\bibfield  {author} {\bibinfo {author} {\bibfnamefont {G.}~\bibnamefont
			{Baym}}, \bibinfo {author} {\bibfnamefont {C.}~\bibnamefont {Pethick}}, \
		and\ \bibinfo {author} {\bibfnamefont {P.}~\bibnamefont {Sutherland}},\
	}\href@noop {} {\bibfield  {journal} {\bibinfo  {journal} {Astrophys. J.}\
		}\textbf {\bibinfo {volume} {170}},\ \bibinfo {pages} {299} (\bibinfo {year}
		{1971})}\BibitemShut {NoStop}%
	\bibitem [{\citenamefont {Carriere}\ \emph {et~al.}(2003)\citenamefont
		{Carriere}, \citenamefont {Horowitz},\ and\ \citenamefont
		{Piekarewicz}}]{Carriere:2002bx}%
	\BibitemOpen
	\bibfield  {author} {\bibinfo {author} {\bibfnamefont {J.}~\bibnamefont
			{Carriere}}, \bibinfo {author} {\bibfnamefont {C.~J.}\ \bibnamefont
			{Horowitz}}, \ and\ \bibinfo {author} {\bibfnamefont {J.}~\bibnamefont
			{Piekarewicz}},\ }\href@noop {} {\bibfield  {journal} {\bibinfo  {journal}
			{Astrophys. J.}\ }\textbf {\bibinfo {volume} {593}},\ \bibinfo {pages} {463}
		(\bibinfo {year} {2003})}\BibitemShut {NoStop}%
	\bibitem [{\citenamefont {Fortin}\ \emph {et~al.}(2016)\citenamefont {Fortin},
		\citenamefont {Providencia}, \citenamefont {Raduta}, \citenamefont
		{Gulminelli}, \citenamefont {Zdunik}, \citenamefont {Haensel},\ and\
		\citenamefont {Bejger}}]{Fortin:2016hny}%
	\BibitemOpen
	\bibfield  {author} {\bibinfo {author} {\bibfnamefont {M.}~\bibnamefont
			{Fortin}}, \bibinfo {author} {\bibfnamefont {C.}~\bibnamefont {Providencia}},
		\bibinfo {author} {\bibfnamefont {A.~R.}\ \bibnamefont {Raduta}}, \bibinfo
		{author} {\bibfnamefont {F.}~\bibnamefont {Gulminelli}}, \bibinfo {author}
		{\bibfnamefont {J.~L.}\ \bibnamefont {Zdunik}}, \bibinfo {author}
		{\bibfnamefont {P.}~\bibnamefont {Haensel}}, \ and\ \bibinfo {author}
		{\bibfnamefont {M.}~\bibnamefont {Bejger}},\ }\href@noop {} {\bibfield
		{journal} {\bibinfo  {journal} {Phys. Rev. C}\ }\textbf {\bibinfo {volume}
			{94}},\ \bibinfo {pages} {035804} (\bibinfo {year} {2016})}\BibitemShut
	{NoStop}%
	\bibitem [{\citenamefont {Abbott~{\sl et al.}}(2018)}]{GW170817_MR_PEsample}%
	\BibitemOpen
	\bibfield  {author} {\bibinfo {author} {\bibfnamefont {B.~P.}\ \bibnamefont
			{Abbott~{\sl et al.}}},\ }\href@noop {} {\bibfield  {journal} {\bibinfo
			{journal} {Phys. Rev. Lett.}\ }\textbf {\bibinfo {volume} {121}},\ \bibinfo
		{pages} {161101} (\bibinfo {year} {2018})}\BibitemShut {NoStop}%
	\bibitem [{\citenamefont {{Abbott}~{\sl et al.}}(2021)}]{GWOSC_softx}%
	\BibitemOpen
	\bibfield  {author} {\bibinfo {author} {\bibfnamefont {R.}~\bibnamefont
			{{Abbott}~{\sl et al.}}},\ }\href@noop {} {\bibfield  {journal} {\bibinfo
			{journal} {SoftwareX}\ }\textbf {\bibinfo {volume} {13}},\ \bibinfo {eid}
		{100658} (\bibinfo {year} {2021})}\BibitemShut {NoStop}%
	\bibitem [{\citenamefont {Xie}\ \emph {et~al.}(2022)\citenamefont {Xie},
		\citenamefont {Chen}, \citenamefont {Ma}, \citenamefont {Guo},\ and\
		\citenamefont {Zhu}}]{Xie2022}%
	\BibitemOpen
	\bibfield  {author} {\bibinfo {author} {\bibfnamefont {W.-J.}\ \bibnamefont
			{Xie}}, \bibinfo {author} {\bibfnamefont {J.-L.}\ \bibnamefont {Chen}},
		\bibinfo {author} {\bibfnamefont {Z.-W.}\ \bibnamefont {Ma}}, \bibinfo
		{author} {\bibfnamefont {J.-H.}\ \bibnamefont {Guo}}, \ and\ \bibinfo
		{author} {\bibfnamefont {L.}~\bibnamefont {Zhu}},\ }\href
	{http://iopscience.iop.org/article/10.1088/1674-1137/ac9888} {\bibfield
		{journal} {\bibinfo  {journal} {Chinese Physics C}\ }\textbf {\bibinfo
			{volume} {47}},\ \bibinfo {pages} {014103} (\bibinfo {year}
		{2022})}\BibitemShut {NoStop}%
	\bibitem [{\citenamefont {Biswas}(2021)}]{Biswas:2021yge}%
	\BibitemOpen
	\bibfield  {author} {\bibinfo {author} {\bibfnamefont {B.}~\bibnamefont
			{Biswas}},\ }\href {\doibase 10.3847/1538-4357/ac1c72} {\bibfield  {journal}
		{\bibinfo  {journal} {Astrophys. J.}\ }\textbf {\bibinfo {volume} {921}},\
		\bibinfo {pages} {63} (\bibinfo {year} {2021})}\BibitemShut {NoStop}%
	\bibitem [{\citenamefont {Grill}\ \emph {et~al.}(2014)\citenamefont {Grill},
		\citenamefont {Pais}, \citenamefont {Provid\^encia}, \citenamefont
		{Vida\~na},\ and\ \citenamefont {Avancini}}]{Grill:2014aea}%
	\BibitemOpen
	\bibfield  {author} {\bibinfo {author} {\bibfnamefont {F.}~\bibnamefont
			{Grill}}, \bibinfo {author} {\bibfnamefont {H.}~\bibnamefont {Pais}},
		\bibinfo {author} {\bibfnamefont {C.}~\bibnamefont {Provid\^encia}}, \bibinfo
		{author} {\bibfnamefont {I.}~\bibnamefont {Vida\~na}}, \ and\ \bibinfo
		{author} {\bibfnamefont {S.~S.}\ \bibnamefont {Avancini}},\ }\href {\doibase
		10.1103/PhysRevC.90.045803} {\bibfield  {journal} {\bibinfo  {journal} {Phys.
				Rev. C}\ }\textbf {\bibinfo {volume} {90}},\ \bibinfo {pages} {045803}
		(\bibinfo {year} {2014})}\BibitemShut {NoStop}%
	\bibitem [{\citenamefont {Pais}\ and\ \citenamefont
		{Provid\^encia}(2016)}]{Pais:2016xiu}%
	\BibitemOpen
	\bibfield  {author} {\bibinfo {author} {\bibfnamefont {H.}~\bibnamefont
			{Pais}}\ and\ \bibinfo {author} {\bibfnamefont {C.}~\bibnamefont
			{Provid\^encia}},\ }\href {\doibase 10.1103/PhysRevC.94.015808} {\bibfield
		{journal} {\bibinfo  {journal} {Phys. Rev. C}\ }\textbf {\bibinfo {volume}
			{94}},\ \bibinfo {pages} {015808} (\bibinfo {year} {2016})}\BibitemShut
	{NoStop}%
	\bibitem [{\citenamefont {Boukari}\ \emph {et~al.}(2021)\citenamefont
		{Boukari}, \citenamefont {Pais}, \citenamefont {Anti\'c},\ and\ \citenamefont
		{Provid\^encia}}]{Boukari:2020iut}%
	\BibitemOpen
	\bibfield  {author} {\bibinfo {author} {\bibfnamefont {O.}~\bibnamefont
			{Boukari}}, \bibinfo {author} {\bibfnamefont {H.}~\bibnamefont {Pais}},
		\bibinfo {author} {\bibfnamefont {S.}~\bibnamefont {Anti\'c}}, \ and\
		\bibinfo {author} {\bibfnamefont {C.}~\bibnamefont {Provid\^encia}},\ }\href
	{\doibase 10.1103/PhysRevC.103.055804} {\bibfield  {journal} {\bibinfo
			{journal} {Phys. Rev. C}\ }\textbf {\bibinfo {volume} {103}},\ \bibinfo
		{pages} {055804} (\bibinfo {year} {2021})}\BibitemShut {NoStop}%
	\bibitem [{Com()}]{CompOSE}%
	\BibitemOpen
	\href {https://compose.obspm.fr} {\enquote {\bibinfo {title} {Compose,
				compstar online supernovae equations of state},}\ }\BibitemShut {NoStop}%
	\bibitem [{\citenamefont {Typel}\ \emph {et~al.}(2022)\citenamefont {Typel}
		\emph {et~al.}}]{CompOSECoreTeam:2022ddl}%
	\BibitemOpen
	\bibfield  {author} {\bibinfo {author} {\bibfnamefont {S.}~\bibnamefont
			{Typel}} \emph {et~al.} (\bibinfo {collaboration} {CompOSE Core Team}),\
	}\href {\doibase 10.1140/epja/s10050-022-00847-y} {\bibfield  {journal}
		{\bibinfo  {journal} {Eur. Phys. J. A}\ }\textbf {\bibinfo {volume} {58}},\
		\bibinfo {pages} {221} (\bibinfo {year} {2022})}\BibitemShut {NoStop}%
	\bibitem [{\citenamefont {{Romero-Shaw}~{\sl et al.}}(2020)}]{Bilby_ref}%
	\BibitemOpen
	\bibfield  {author} {\bibinfo {author} {\bibfnamefont {I.~M.}\ \bibnamefont
			{{Romero-Shaw}~{\sl et al.}}},\ }\href@noop {} {\bibfield  {journal}
		{\bibinfo  {journal} {Mon. Not. Roy. Astron. Soc.}\ }\textbf {\bibinfo
			{volume} {499}},\ \bibinfo {pages} {3295} (\bibinfo {year}
		{2020})}\BibitemShut {NoStop}%
\end{thebibliography}
	
\end{document}